\documentclass[twocolumn,english,aps,prb,twocolum,superscriptaddress,bibnotes,amsmath,amssymb,floatfix]{revtex4-1}
\usepackage[colorlinks=true,citecolor=blue,linkcolor=magenta]{hyperref}

\usepackage[markup=nocolor, authormarkupposition=left]{changes} 
\usepackage{soul}
\usepackage[utf8]{inputenc}
\usepackage[english]{babel}
\usepackage{amsmath,amsfonts,amssymb}
\usepackage[T1]{fontenc}
\usepackage{url}

\usepackage{amsmath}
\usepackage{amsfonts}
\usepackage{amssymb}

\usepackage{epstopdf}
\usepackage{graphicx}
\graphicspath{{./Figures/}}

\usepackage{changes}

\begin{document}

\title{A chip-integrated comb-based microwave oscillator}

\author{Wei Sun}
\thanks{These authors contributed equally to this work.}
\affiliation{International Quantum Academy, Shenzhen 518048, China}

\author{Zhiyang Chen}
\thanks{These authors contributed equally to this work.}
\affiliation{International Quantum Academy, Shenzhen 518048, China}
\affiliation{Shenzhen Institute for Quantum Science and Engineering, Southern University of Science and Technology, Shenzhen 518055, China}

\author{Linze Li}
\thanks{These authors contributed equally to this work.}
\affiliation{School of Information Science and Technology, ShanghaiTech University, Shanghai 201210, China}

\author{Chen Shen}
\thanks{These authors contributed equally to this work.}
\affiliation{International Quantum Academy, Shenzhen 518048, China}
\affiliation{Qaleido Photonics, Hangzhou 310000, China}

\author{Jinbao Long}
\affiliation{International Quantum Academy, Shenzhen 518048, China}

\author{Huamin Zheng}
\affiliation{International Quantum Academy, Shenzhen 518048, China}
\affiliation{College of Electronics and Information Engineering, Shenzhen University, Shenzhen 518000, China}

\author{Luyu Wang}
\affiliation{School of Information Science and Technology, ShanghaiTech University, Shanghai 201210, China}

\author{Qiushi Chen}
\affiliation{School of Information Science and Technology, ShanghaiTech University, Shanghai 201210, China}

\author{Zhouze Zhang}
\affiliation{School of Information Science and Technology, ShanghaiTech University, Shanghai 201210, China}

\author{Baoqi Shi}
\affiliation{International Quantum Academy, Shenzhen 518048, China}
\affiliation{Department of Optics and Optical Engineering, University of Science and Technology of China, Hefei 230026, China}

\author{Shichang Li}
\affiliation{International Quantum Academy, Shenzhen 518048, China}
\affiliation{Shenzhen Institute for Quantum Science and Engineering, Southern University of Science and Technology,
Shenzhen 518055, China}

\author{Lan Gao}
\affiliation{International Quantum Academy, Shenzhen 518048, China}
\affiliation{Shenzhen Institute for Quantum Science and Engineering, Southern University of Science and Technology,
Shenzhen 518055, China}

\author{Yi-Han Luo}
\affiliation{International Quantum Academy, Shenzhen 518048, China}

\author{Baile Chen}
\email[]{chenbl@shanghaitech.edu.cn}
\affiliation{School of Information Science and Technology, ShanghaiTech University, Shanghai 201210, China}

\author{Junqiu Liu}
\email[]{liujq@iqasz.cn}
\affiliation{International Quantum Academy, Shenzhen 518048, China}
\affiliation{Hefei National Laboratory, University of Science and Technology of China, Hefei 230088, China}

\maketitle

\noindent\textbf{Low-noise microwave oscillators are cornerstones for wireless communication, radar and clocks. 
The employment and optimization of optical frequency combs have enabled photonic microwave synthesizers with unrivalled noise performance and bandwidth breaking the bottleneck of those electronic counterparts \cite{Fortier:11, Xie:17, Li:14, Nakamura:20}. 
Emerging interest has been concentrated on photonic microwave generation using chip-based Kerr frequency combs, namely microcombs \cite{Kippenberg:18, Pasquazi:18}. 
Today microcombs built on photonic integrated circuits feature small size, weight and power consumption \cite{Moss:13, Gaeta:19, Liu:21}, and can be designed and manufactured to oscillate at any target frequency ranging from microwave to millimeter-wave band \cite{Liang:15, Liu:20, Yao:22, Tetsumoto:21, Wang:21}. 
However, a monolithic microcomb-based microwave oscillator requires photonic integration of lasers, photodetectors and nonlinear microresonators on a common substrate, which has still remained elusive. 
Here, we demonstrate the first of such a fully integrated, microcomb-based, photonic microwave oscillator at 10.7 GHz. 
The chip device, powered by a customized microelectronic circuit, leverages hybrid integration of a high-power DFB laser, a Si$_3$N$_4$ microresonator of a quality factor exceeding 25$\times$10$^6$, and a high-speed photodetector chip of 110 GHz bandwidth (3-dB) and 0.3 A/W responsivity.
Each component represents the state of the art of its own class, yet also allows large-volume manufacturing with low cost using established CMOS foundries. 
The hybrid chip outputs an ultralow-noise laser of 6.9 Hz intrinsic linewidth, a coherent microcomb of 10.7 GHz repetition rate, and a 10.7 GHz microwave carrier of 6.3 mHz linewidth -- all the three functions in one entity occupying a footprint of only 76 mm$^2$.
The microwave phase noise reaches $-$75/$-$105/$-$130 dBc/Hz at 1/10/100 kHz Fourier offset frequency. 
Our results can reinvigorate our information society for communication, sensing, imaging, timing and precision measurement.
}

\noindent\textbf{Introduction}. 
Low-noise microwave oscillators are ubiquitously deployed in our information society for communication, sensing, spectroscopy and timing. 
A particularly important frequency band is the microwave X-band (8 to 12 GHz), dedicated for radar, wireless networks, and satellite communication. 
Currently, a paradigm shift is ongoing which utilizes photonics to synthesize X-band microwaves \cite{Fortier:11, Xie:17, Li:14, Liang:15, Liu:20, Yao:22, LiJ:13, LiJ:23, Tang:18}.
Compared to the electronic counterpart, the 
photonic microwave generation enables unrivalled performance, particularly in noise and bandwidth. 

Among all photonics-based approaches, optical frequency combs \cite{Cundiff:03, Fortier:19, Diddams:20}, which coherently link radio- or microwave frequency to optical frequency, have enabled microwaves with unrivalled spectral purity (noise) by leveraging optical frequency division (OFD) \cite{Fortier:11, Xie:17, Li:14}. 
Optical-clock-referenced OFD \cite{Nakamura:20} has created 10 GHz microwaves with $10^{-16}$ fractional frequency instability at 1 s, two orders of magnitude more stable than the cesium fountain clocks defining the SI second. 
Currently, by virtue of high-$Q$ optical microresonators and harnessing the Kerr nonlinearity, an emerging paradigm is to develop and apply microresonator-based Kerr frequency combs, i.e. ``microcombs'' \cite{Kippenberg:18, Pasquazi:18, Herr:14, Yi:15, Brasch:15, Joshi:16, Xue:15} that have small size, weight and power consumption. 
In particular, exploiting recent breakthroughs in fabrication of ultralow-loss photonic integrated circuits (PIC) \cite{Moss:13, Gaeta:19, Liu:21}, optical microresonators can now be constructed on a plethora of integrated material platforms \cite{Kovach:20, Chang:22}. 
These PIC-based microresonators have permitted chip-level microcombs with diverse merits and functions. 
Notably, latest endeavors have demonstrated microcomb-based OFD \cite{Kudelin:23, Sun:23, Jin:24, He:24}.
Despite, these OFD setups are still bulky, and require sophisticated locking techniques and external references. 
Therefore, a free-running microcomb-based microwave oscillator that is fully integrated at chip scale is still beneficial, though the phase noise can be compromised. 

%%%%%%%%%%%%%%%%%%%%%%%%%%%%%%%%%%%%%%%%%%%
\begin{figure*}[t!]
\centering
\includegraphics{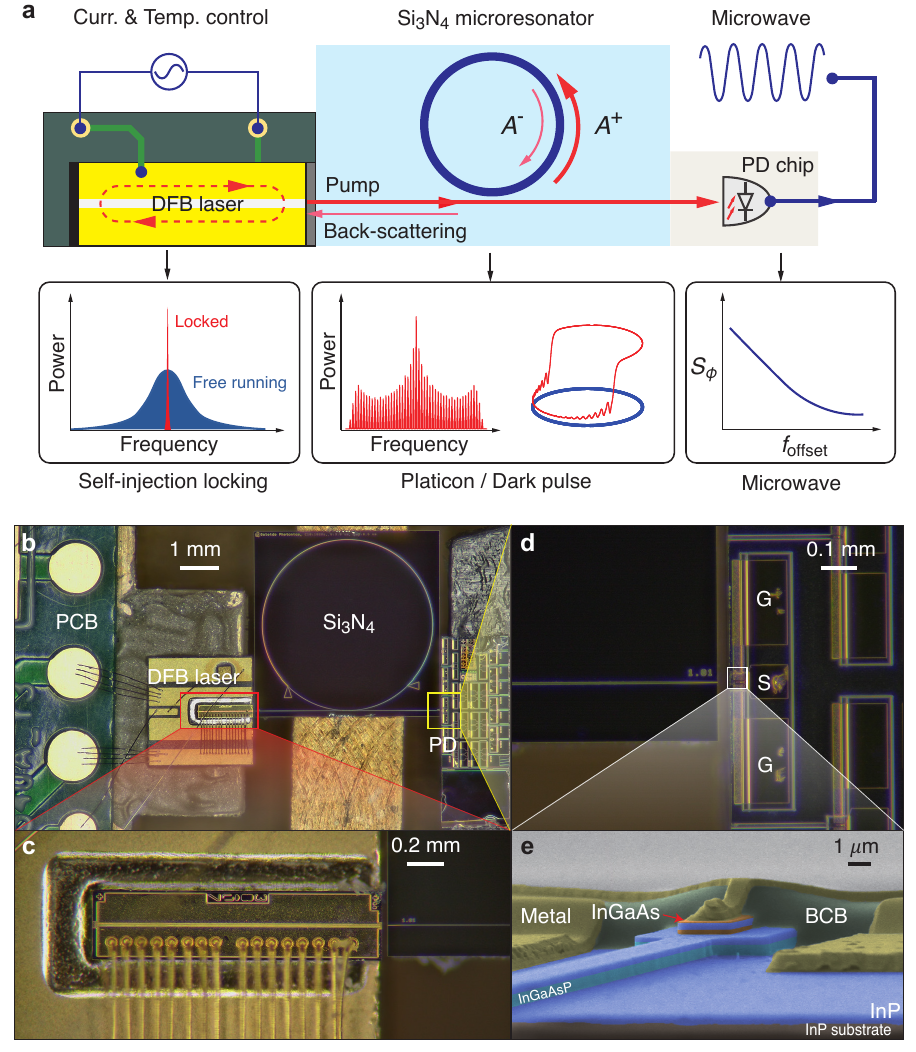}
\caption{
\textbf{Schematic and images of the hybrid, microcomb-based, photonic microwave oscillator}.
\textbf{a}. 
Schematic of the chip device. 
The microelectronic circuit supplies current to the DFB laser and stabilizes its temperature. 
Via edge-coupling, the CW light from the laser enters the Si$_3$N$_4$ microresonator ($A^+$). 
Exploiting the optical back-scattering ($A^-$) from the Si$_3$N$_4$ microresonator to the laser, laser self-injection locking occurs that significantly narrows the laser linewidth. 
Simultaneously a circulating platicon/dark-pulse stream is formed in the microresonator. 
The output pulse stream is received by the PD chip that outputs a microwave carrier at the pulse repetition rate.
$S_{\phi}$, microwave phase noise.
$f_\text{offset}$, Fourier offset frequency.
\textbf{b}.
Photo of the hybrid chip device and individual components.  
\textbf{c}.
Zoom-in image showing the DFB laser wire-bonded to a PCB and edge-coupled to the Si$_3$N$_4$ microresonator chip. 
\textbf{d}.
Zoom-in image showing the the PD chip edge-coupled to the Si$_3$N$_4$ chip and connected by a ground-source-ground (GSG) probe for microwave signal output. 
\textbf{e}.
False-colored SEM image showing PD's multilayer structure. 
InGaAs, indium gallium arsenide.
InGaAsP, Indium gallium arsenide phosphide.
InP, Indium phosphide.
BCB, benzocyclobutene, a kind of resin.
}
\vspace{5cm}
\label{Fig:1}
\end{figure*}
%%%%%%%%%%%%%%%%%%%%%%%%%%%%%%%%%%%%%%%%%%%

%%%%%%%%%%%%%%%%%%%%%%%
Here, we demonstrate the first fully integrated, microcomb-based, photonic microwave oscillator. 
The chip device leverages hybrid integration \cite{Stern:18, Raja:19, Shen:20} of a distributed-feedback (DFB) laser chip, a high-$Q$ silicon nitride (Si$_3$N$_4$) microresonator chip, and a photodetector (PD) chip. 
The schematic and components are depicted in Fig. \ref{Fig:1}.
The DFB laser chip is driven and stabilized by a microelectronic circuit, and emits CW light at 1550 nm. 
Via edge-coupling, the light enters the Si$_3$N$_4$ microresonator of 10.7 GHz free spectral range (FSR), where a coherent platicon microcomb of 10.7 GHz repetition rate is formed. 
The output light from the Si$_3$N$_4$ microresonator is delivered to the PD chip.
Upon photodetection of the microcomb's repetition rate \cite{Liang:15, Liu:20, Yao:22}, the PD outputs a microwave carrier of 10.7 GHz and its harmonics. 
The entire device occupies a footprint of only 76 mm$^2$.

%%%%%%%%%%%%%%%%%%%%%%%
\section*{Characterization of individual components}
\noindent\textbf{Laser}. 
We use a commercial DFB laser as shown in Fig. \ref{Fig:1}b,c. 
The laser outputs 160 mW CW light at 1549.9 nm with 500 mA laser current. 
The free-running DFB laser exhibits 55 mA current at laser threshold, transverse-electric (TE) polarization, and 1 nm wavelength tunability over a laser current up to 500 mA. 
The laser's temperature is stabilized at 30$^\circ$C using a thermo-electric cooler (TEC).
A printed circuit board (PCB) provides driving current and temperature stabilization to the laser chip.
The laser chip is edge-coupled to the Si$_3$N$_4$ chip \cite{Raja:19, Shen:20, Lihachev:22a}. 
The laser output waveguide is aligned to the inverse taper of the Si$_3$N$_4$ bus waveguide and side-coupled to the microresonator. % \cite{Pfeiffer:17b}.  
The coupling efficiency between the laser chip and the Si$_3$N$_4$ chip is measured as 23\%, i.e. with 6.4 dB insertion loss. 
Extra characterization data are shown in Supplementary Information. 
This loss value can be reduced in the future using optimized inverse taper that matches the laser output mode.

%%%%%%%%%%%%%%%%%%%%%%%%%%%%%%%%%
\noindent\textbf{Microresonator}.
The Si$_3$N$_4$ chip contains a high-$Q$ microresonator of 10.7 GHz FSR, as shown in Fig. \ref{Fig:1}b. 
A high $Q$ is critical for microcomb generation, since the power threshold $P_0$ for Kerr parametric oscillation is $P_0\propto 1/(D_1Q^2)$, where $D_1/2\pi$ is the FSR \cite{Kippenberg:18, Pasquazi:18}. 
We fabricate Si$_3$N$_4$ waveguides of 300 nm thickness using a deep-ultraviolet (DUV) subtractive process on 150-mm-diameter (6-inch) wafers \cite{Ye:23}. 
The optimized fabrication process is described in Supplementary Information. 
We choose 300 nm Si$_3$N$_4$ thickness for the following reasons. 
First, since high-quality Si$_3$N$_4$ films deposited via low-pressure chemical vapor deposition (LPCVD, as used in our case) are prone to crack due to intrinsic tensile stress \cite{Luke:13}, Si$_3$N$_4$ films with thickness below 300 nm are free from cracks. 
Second, 300-nm-thick Si$_3$N$_4$ fabrication process is currently established as a standard process in nearly all CMOS foundries worldwide \cite{Munoz:19, Xiang:22a}. 

\begin{figure*}[t!]
\centering
\includegraphics{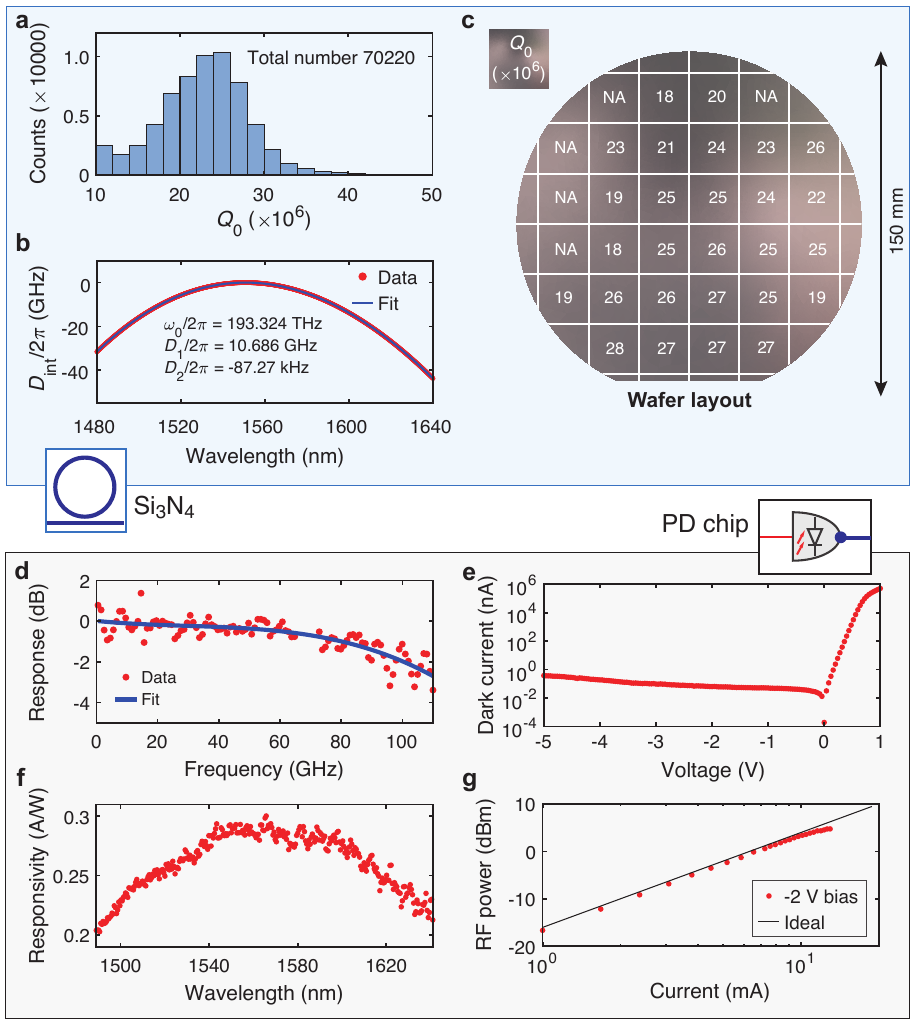}
\caption{
\textbf{Characterization of Si$_3$N$_4$ microresonators and the photodetector chip}. 
\textbf{a}. 
Histogram of 70,220 measured intrinsic quality factor $Q_0$ from twenty-seven Si$_3$N$_4$ chips on a 6-inch wafer. 
The most probable value is $Q_0=25\times10^6$.
\textbf{b}. 
Profile of microresonator's integrated dispersion $D_\text{int}/2\pi$ of the 10.7-GHz-FSR Si$_3$N$_4$ microresonator.
Red dots are measured data, and the blue line is the polynomial fit. % of the measured data.
$\omega_0/2\pi=193.324$ THz is the reference frequency corresponding to the pump DFB's frequency.
$D_1/2\pi=10.686$ GHz is the microresonator FSR.
$D_2/2\pi=-87.27$ kHz is the normal GVD. 
See Method for the definition of $D_\text{int}$.
\textbf{c}. 
The most probable values $Q_0$ of the Si$_3$N$_4$ chips at different places uniformly across the 6-inch wafer. 
In most places $Q_0\geq 20\times 10^6$ values are found, showing high yield of the Si$_3$N$_4$ fabrication process. 
NA, not available.
\textbf{d}. 
Frequency response of the $3\times15$ $\mu \text{m}^2$ PD chip. 
Red dots are measured data, and the blue line is the polynomial fit.
\textbf{e}. 
Measured dark current versus bias voltage of the PD chip. 
Negative bias voltage leads to dark current below 1 nA.
\textbf{f}. 
Measured responsivity versus wavelength of the PD chip.
\textbf{g}. 
RF power versus the AC current of the PD chip. 
Measured data (red dots) is comparable with the ideal case (black line).
}
\label{Fig:2}
\end{figure*}

We characterize 70,220 resonances in the fundamental transverse-electric (TE$_{00}$) mode from 27 chips uniformly distributed over the 6-inch wafer, using a vector spectrum analyzer \cite{Luo:23} covering 1480 to 1640 nm (see Method). 
The waveguide width of the Si$_3$N$_4$ microresonator is 2.6 $\mu$m. 
Each resonance's intrinsic linewidth $\kappa_0/2\pi$ and central frequency $\omega/2\pi$ are measured and fitted.
The histogram of intrinsic quality factors $Q_0=\omega/\kappa_0$ of the 70,220 resonances is shown in Fig. \ref{Fig:2}a. 
The most probable value is $Q_0=25\times 10^6$, corresponding to a linear optical loss of $\alpha\approx1.3$ dB/m (physical length).
Figure \ref{Fig:2}c shows the measured $Q_0$ distribution over the entire wafer, where $Q_0\geq 20\times 10^6$ is found in most places. 
Such a high $Q$ value is sufficient for coherent microcomb generation in the Si$_3$N$_4$ microresonator with 37.7 mW power in the bus waveguide. 

There are two types of coherent microcombs, i.e. the bright dissipative Kerr soliton (DKS) \cite{Herr:14, Yi:15, Brasch:15, Joshi:16, Liang:15}, and the dark pulse or platicon \cite{Xue:15, Lobanov:15, Huang:15b, Parra-Rivas:16, Nazemosadat:21}. 
Figure \ref{Fig:2}b shows the measured microresonator's integrated dispersion $D_\text{int}$ (see definition in Method). 
Our Si$_3$N$_4$ microresonator features normal group velocity dispersion (GVD, $D_2<0$), permitting platicon generation. 
In contrast, DKS requires anomalous GVD ($D_2>0$) that necessitates Si$_3$N$_4$ thickness above 600 nm and specific dispersion engineering \cite{Okawachi:14, Luke:13}. 

In addition, we emphasize another two advantages of using platicons instead of DKS for microwave generation. 
First, compared to DKS, platicons exhibit remarkably higher CW-to-pulse power conversion efficiency \cite{Xue:17b, JangJ:21}. 
For 10.7 GHz repetition rate, we achieve $8\%$ power conversion efficiency using platicons (discussed later), while the typical value is $\sim0.2\%$ for DKS \cite{Liu:20, JangJ:21}. 
Second, for DKS-based microwave generation, the CW pump must be filtered or blocked before photodetection of the pulse's repetition rate \cite{Liang:15, Liu:20, Yao:22}. 
Otherwise, the overwhelmingly strong pump can saturate the PD, yielding deteriorated power and phase noise of the microwave signal. 
Therefore, a filtering element, e.g. a notch filter, is required in the DKS-based microwave oscillator, which however complicates the overall system and photonic integration. 
In contrast, platicon-based microwave oscillator does not require such a filter, as in our case. 
In fact, the presence of the strong CW pump is beneficial. 
The pump beats against its two neighboring comb lines -- with the highest power among all comb lines -- and generates strong microwave signal at the repetition rate. 
In sum, we employ platicons for photonic microwave generation. 

%%%%%%%%%%%%%%%%%%%%%%%%%%%%%%%%%%%%%%%%%%%
%%%%%%%%%%%%%%%%%%%%%%%%%%%%%%%%%%%%%%%%%%%
%%%%%%%%%%%%%%%%%%%%%%%%%%%%%%%%%
\noindent\textbf{Photodetector}.
The high-speed PD is edge-coupled to the Si$_3$N$_4$ chip's output waveguide, and is connected by a ground-source-ground (GSG) probe for microwave signal output, as shown in Fig. \ref{Fig:1}d. 
The epitaxial structure of the PD chip is grown on a semi-insulating indium phosphide (InP) substrate \cite{Li:23}.
Figure \ref{Fig:1}e shows the false-colored scanning electron microscope (SEM) image of the PD’s cross-section and multi-layer structure.
The fabrication process of the PD chip starts with P-type contact metals (Ti/Pt/Au/Ti) deposition. 
Dry etching steps are then performed using inductively coupled plasma etching to form a triple-mesa structure. 
After the deposition of N-type contact metals (GeAu/Ni/Au), a benzocyclobutene (BCB) layer is implemented beneath the coplanar waveguides (CPWs). 
This approach eliminates the necessity for air-bridge structures, thus ensuring a consistent and stable connection between p-mesa and CPWs. 
Details of the fabrication process are found in Supplementary Information.

The frequency response of the InGaAs/InP waveguide device is investigated by an optical heterodyne setup. 
Upon receiving a heterodyne beatnote between two tunable lasers with wavelengths near 1550 nm, the PD outputs a high-frequency RF signal whose power is measured by a Rohde \& Schwarz powermeter (NRP-Z58). 
Details of the characterization setup are found in Ref. \cite{Li:23}. 
A typical frequency response for $3\times15$ $\mu$m$^2$ PD chip is shown in Fig.~\ref{Fig:2}d. 
By a polynomial fit of the measured data, 3-dB bandwidth over 110 GHz is shown.
The I-V characteristics of the $3\times15$ $\mu$m$^2$ PD chip are measured by a semiconductor device analyzer. 
Figure~\ref{Fig:2}e shows the measured dark current below 1 nA, which out-performs commercial high-speed PDs.
Figure \ref{Fig:2}f shows the responsivity as high as 0.3 A/W and above 0.2 A/W within 152 nm bandwidth.  
Figure \ref{Fig:2}g shows the voltage-dependent saturation property of the PD measured around 10 GHz.
The ideal relationship between RF power $P_\text{RF}$ and DC photo-current $I_\text{dc}$ is $P_\text{RF}= I_\text{dc}^2R_\text{load}/2$, assuming a 100\% modulation depth ($I_\text{ac}=I_\text{dc}$), where $R_\text{load}=50\ \Omega$ is the load resistance and $I_{\text{ac}}$ is AC photo-current.

%%%%%%%%%%%%%%%%%%%%%%%%%%%%%%%%%%%%%%%%%%%
\begin{figure*}[t!]
\centering
\includegraphics[width=14.5cm]{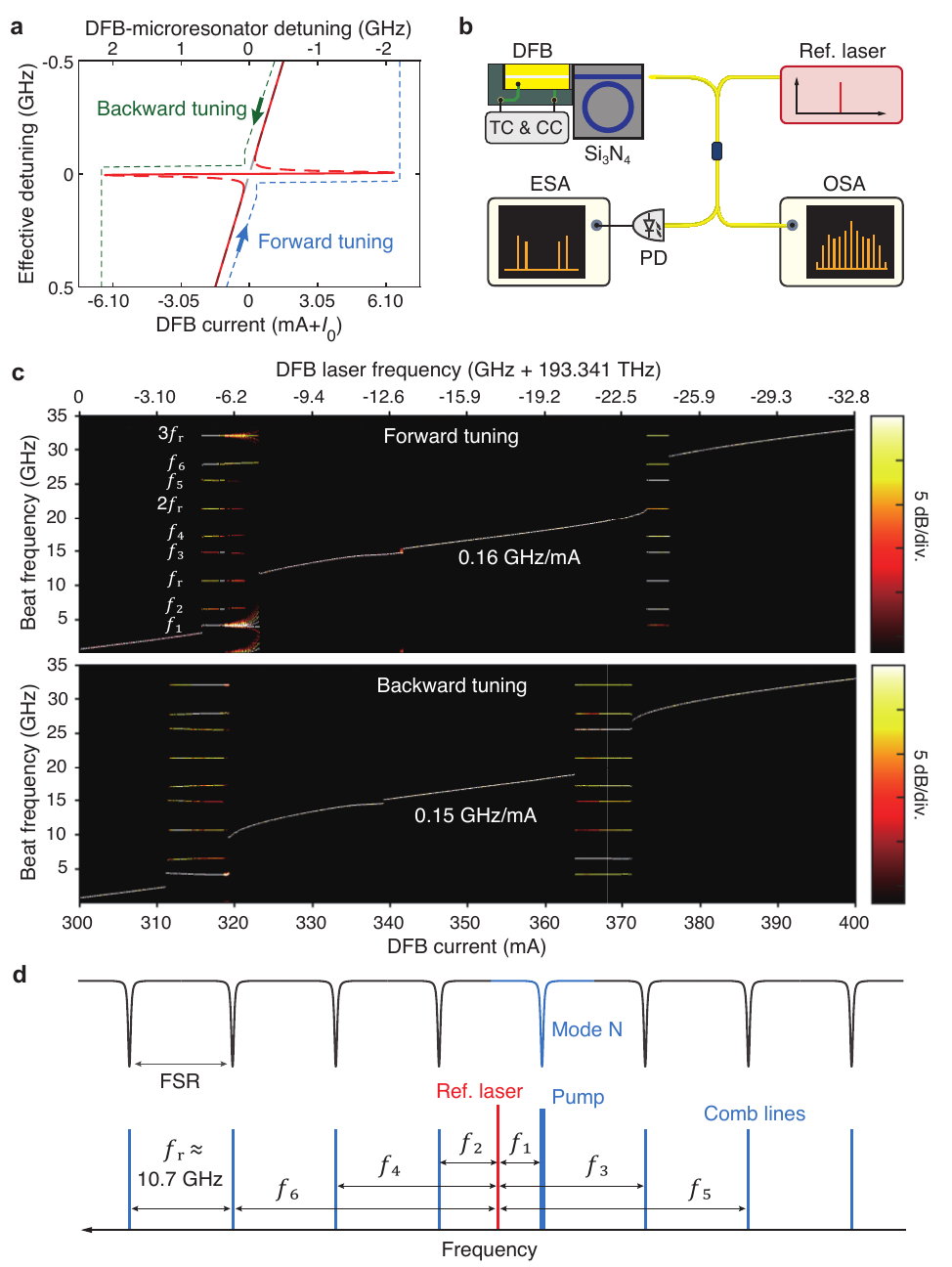}
\caption{
\textbf{Characterization of laser self-injection locking.}
\textbf{a}. 
Theoretical model of SIL dynamics using our experimental parameters. 
The red/gray line represents the steady-state solution with/without SIL.
The vertical axis is the frequency detuning between the SIL DFB laser to the microresonator's resonance.
The top horizontal axis is the frequency detuning between the free-running DFB laser to the microresonator's resonance. 
The bottom horizontal axis is the DFB current relative to $I_0$. 
The blue/green dashed line marks the forward/backward tuning curves. 
\textbf{b}. 
Experimental setup. 
OSA, optical spectrum analyser.
ESA, electronic spectrum analyser.
TC/CC, temperature/current control.
\textbf{c}. 
Measured SIL dynamics. 
The vertical axis is the beat frequency $f$ between the reference laser and the SIL laser.
In the forward/backward tuning, the DFB current increases/decreases and thus the laser frequency decreases/increases. 
Color bar marks the beat signal's power.
The flat segments signal the occurrence of SIL. 
The multiple beat frequencies ($f_1$ to $f_6$ and $f_r$ to $3f_r$) signal platicon microcomb generation.
\textbf{d}.
Illustration of the reasons for $f_1$ to $f_6$ and $f_r$ to $3f_r$. 
Upper panel shows the microresonator's resonance grid.
Lower panel illustrates that, the microcomb's lines (blue) beat with the reference laser (red) and generate $f_1$ to $f_6$; 
the microcomb's lines beat among themselves and generate $f_r$, $2f_r$ and $3f_r$. 
}
\label{Fig:3}
\end{figure*}
%%%%%%%%%%%%%%%%%%%%%%%%%%%%%%%%%%%%%%%%%%%

%%%%%%%%%%%%%%%%%%%%%%%%%%%%%%%%%%%%%%%%%%%
\begin{figure*}[t!]
\centering
\includegraphics{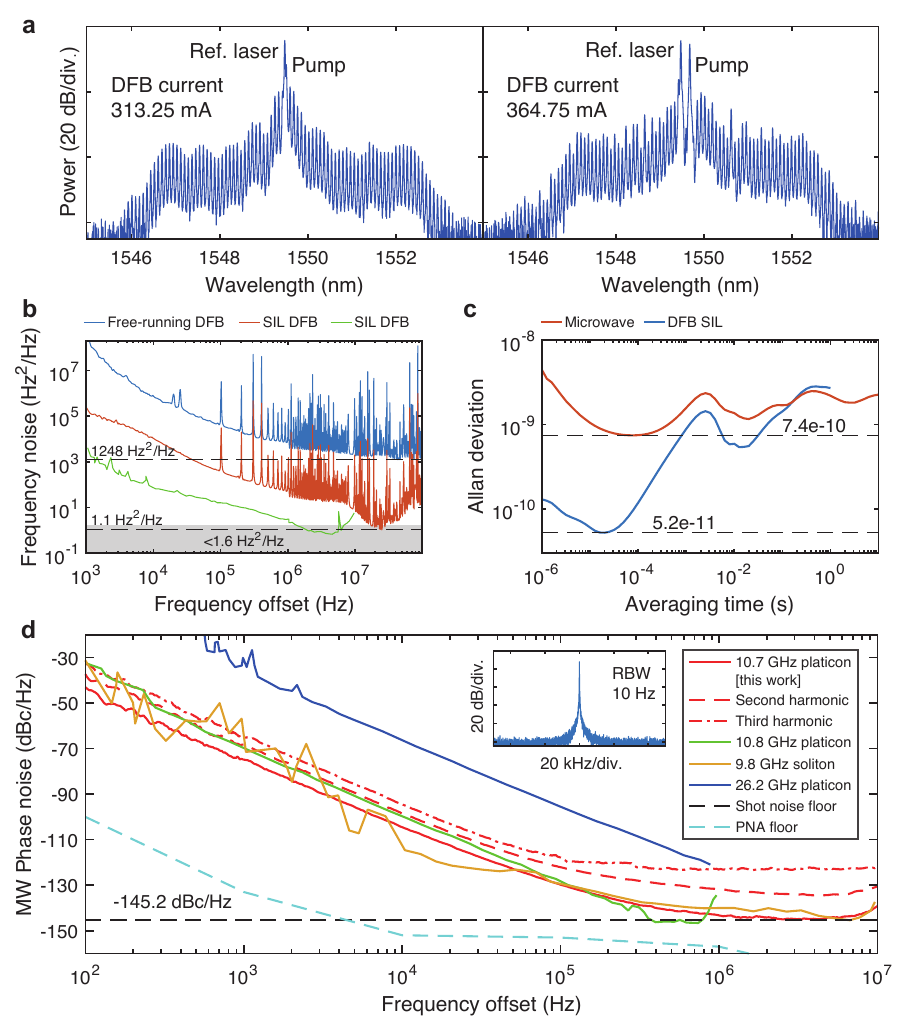}
\caption{
\textbf{Characterization of platicon microcombs and microwaves.}
\textbf{a}. 
Optical spectra of the generated platicons with DFB current of 313.25 and 364.75 mA -- the two SIL regions in the backward tuning in Fig. \ref{Fig:3}c.
\textbf{b}. 
Frequency noise spectra of the free-running (blue) and SIL (red) DFB lasers.
The frequency noise of the free-running DFB laser is 1248 Hz$^2$/Hz corresponding to 7.8 kHz intrinsic linewidth.
The SIL laser's frequency noise reaches 1.1 Hz$^2$/Hz corresponding to 6.9 Hz intrinsic linewidth.
Gray area marks the region below 1.6 Hz$^2$/Hz corresponding to 10 Hz intrinsic linewidth.
Green data is the result in Ref.\cite{Jin:21}.
\textbf{c}. 
Measured Allan deviations of the SIL laser (red) and the generated 10.7 GHz microwave (blue). 
\textbf{d}. 
Phase noise spectra of our 10.7 GHz microwave (solid red line), the second harmonic (21.4 Ghz, dashed red line), and the third harmonic (32.1 GHz, dashdot red line), in comparison with the spectra in Ref.\cite{Jin:21} (green line), Ref.\cite{Liu:20} (orange line), and Ref.\cite{Lihachev:22a} (blue line), as well as the PNA floor (light dashed blue line) and shot noise floor (dashed black line).
While we use an integrated PD chip, Ref.\cite{Jin:21, Liu:20, Lihachev:22a} use commercial, non-integrated PD via fiber connection. 
Inset is the power spectrum of the 10.7 GHz microwave in our work.
RBW, resolution bandwidth.
}
\label{Fig:4}
\end{figure*}
%%%%%%%%%%%%%%%%%%%%%%%%%%%%%%%%%%%%%%%%%%%

%%%%%%%%%%%%%%%%%%%%%%%
\section*{Experimental results}
\noindent\textbf{Narrow-linewidth laser and platicon microcomb}.
First, we characterize the laser dynamics and platicon states with the PD chip removed.  
The free-running DFB laser's frequency is increased/decreased by decreasing/increasing the laser driving current. 
By scanning the DFB current in the forward/backward direction (increasing/decreasing the current) between 300 and 400 mA, the laser frequency is tuned by 32.8 GHz, i.e. with tuning coefficient of $-0.328$ GHz/mA. 
When the laser frequency approaches a microresonator resonance, light $A^+$ is coupled into the Si$_3$N$_4$ microresonator, as shown in Fig. \ref{Fig:1}a. 
Due to surface roughness and bulk inhomogeneity of the Si$_3$N$_4$ waveguide, back-scattered light $A^-$ is generated within the microresonator, and is injected back into the laser without an optical isolator. 
This back-scattered light can trigger laser self-injection locking (SIL) \cite{Liang:15b, Kondratiev:17, Kondratiev:23, Voloshin:21}, which locks the laser frequency $\nu$ to the microresonator resonance, as shown in Fig. \ref{Fig:3}a (see Supplementary Information). 
The locking range $\Delta \nu_\text{lock}$ is defined as \cite{Kondratiev:17, Kondratiev:23} 
\begin{equation}
    \Delta \nu_\text{lock} \approx 2\nu\sqrt{1+\alpha_\text{g}^2}\frac{\eta \beta}{Q_\text{d}R_\text{o}}.
\end{equation}
where $\alpha_\text{g}$ is the phase-amplitude factor of the DFB laser, 
$\eta=\kappa_\text{ex}/(\kappa_\text{ex}+\kappa_0)$ is the coupling efficiency with $\kappa_\text{ex}$ being the external coupling rate and $\kappa_0$ being the intrinsic loss,  
$\beta$ is the coupling rate between the counter-propagating modes ($A^+$ and $A^-$) in unit of $\kappa/2$ with $\kappa = \kappa_\text{ex}+\kappa_0$, 
$Q_\text{d}$ is the DFB laser cavity's quality factor, $R_\text{o}$ is the reflectivity of the output facet of the DFB laser.

Once SIL occurs, the intrinsic laser linewidth is significantly suppressed due to the high quality factor $Q_\text{m}$ of the Si$_3$N$_4$ microresonator which is much higher than $Q_\text{d}$ \cite{Liang:15b, Kondratiev:17, Kondratiev:23}. 
The suppression ratio of intrinsic laser linewidth can be expresses as
\begin{equation}
    \frac{\delta \nu_\text{lock}}{\delta \nu_\text{free}} \approx \frac{Q_\text{d}^2}{Q_\text{m}^2}\frac{1}{64\eta^2 \beta^2 (1+\alpha_\text{g}^2) },
\end{equation}
where $\delta \nu_\text{lock}$ is the locked laser's intrinsic linewidth, and $\delta \nu_\text{free}$ is the free-running laser's intrinsic linewidth.
In addition, with sufficient intra-cavity optical power in the Kerr-nonlinear Si$_3$N$_4$ microresonator, nonlinear SIL can lead to DKS \cite{Raja:19, Shen:20, Voloshin:21, Pavlov:18} or platicon formation \cite{Jin:21, Lihachev:22a}, depending on the microresonator GVD. 
In our case with normal GVD, platicons are generated.

The experimental setup to characterize nonlinear SIL dynamics and platicon generation is shown in Fig. \ref{Fig:3}b. 
A tunable diode laser (Toptica CTL 1550) is frequency-stabilized to a self-referenced fiber optical frequency comb (Quantum CTek). 
It serves as a reference laser to beat against the DFB laser emitted from the Si$_3$N$_4$ chip.
The beat signal's frequency $f$ is recorded by an electronic spectrum analyser (ESA, Rohde \& Schwarz FSW43), and the optical spectrum is recorded by an optical spectrum analyser (OSA, Yokogawa AQ6370D).

As shown in Fig. \ref{Fig:3}c, while the beat frequency $f$ has a global monotonic dependence on the DFB current, it also exhibits multiple flat segments that stay nearly constant when the DFB current varies by less than 10 mA.
For example, in the backward tuning, $f$ is nearly constant from $\sim372$ to $\sim363$ mA and from $\sim320$ to $\sim311$ mA.
Such behaviors signal the occurrence of SIL and platicon generation. 
Degrading SIL performance can be resulted from the presence of the Si$_3$N$_4$ bus waveguide that serves as an extra Fabry-P\'erot cavity with weak reflection (see Supplementary information).

The multiple beat frequencies ($f_1$ to $f_6$) in Fig. \ref{Fig:3}c within the SIL region, e.g. around 320 mA DFB current in the forward tuning, are illustrated in Fig.~\ref{Fig:3}d. 
Assuming that the pump laser frequency is close to the resonance mode $N$.
Then $f_1$ to $f_6$ are the beat frequencies between the reference laser and the platicon's comb lines including the pump. 
Meanwhile, $f=m\cdot f_r$ ($m$ is an integer) is the beat frequencies among the comb lines, where $f_r\approx10.7$ GHz is the mode spacing (i.e. the platicon's repetition rate).
As the DFB current is tuned to 372 mA in the forward tuning, the laser frequency approaches the $N+1$ resonance mode, leading to another SIL region. 
Identical behaviors are found in the backward tuning. 
Considering both directions, the DFB current can be tuned by 12.3 mA in the SIL region, corresponding to a full locking range about 4.0 GHz.
Such nonlinear SIL dynamics with platicon generation is similar to the dynamics with DKS generation in Ref. \cite{Voloshin:21}.

The optical spectra of platicon states at 313.25 and 364.75 mA DFB current in the backward tuning are shown in Fig.~\ref{Fig:4}a.
In the SIL region and with 313.25 mA current, the output microcomb power excluding the pump line from the Si$_3$N$_4$ chip is about 1.1 mW.
When the DFB laser frequency is off-resonant, the CW output power from the Si$_3$N$_4$ chip is 14 mW. 
Therefore the CW-to-platicon conversion efficiency is calculated as 8\%.
Figure \ref{Fig:4}b shows the optical frequency noise of the SIL laser characterized by a delayed self-heterodyne method \cite{Yuan:22} (see Supplementary Information).
The lowest frequency noise is 1.1 Hz$^2$/Hz, corresponding to 6.9 Hz intrinsic linewidth and 31-dB noise suppression on the free-running DFB laser.

\noindent\textbf{Low-noise microwave}.
The demonstrated narrow linewidth and ultralow noise of the SIL laser are critical for low-noise microwave generation, since the microwave phase noise inherits the laser phase noise with a certain transduction factor \cite{Liu:20, Tetsumoto:21}. 
Finally, we edge-couple the PD chip to the Si$_3$N$_4$ chip, enabling photonic microwave generation up to 110 GHz within the PD's bandwidth. 
Microwave carriers with frequencies corresponding to the platicon's repetition rate of $f_r\approx10.7$ GHz and harmonics of $f_r$ are output.
Here, we focus on the 10.7 GHz microwave and its second (21.4 GHz in the K-band) and third harmonics (32.1 GHz in the K$_\text{a}$-band). 
The RF power exported from the PD chip is $-28$/$-30$/$-33$ dBm for the 10.7/21.4/32.1 GHz microwave.
The 10.7 GHz microwave's power spectrum is shown in Fig.~\ref{Fig:4}d inset with 6.3 mHz measured linewidth.
The microwave's phase noise is measured with 20-times averaging by a phase noise analyzer (PNA, Rohde \& Schwarz FSWP50), as shown in Fig.~\ref{Fig:4}d. 
The 10.7 GHz microwave's phase noise reaches $-75$/$-105$/$-130$ dBc/Hz at 1/10/100 kHz Fourier offset frequency. 
The phase noise of the second-/third-harmonic microwave is about 6.0/9.5 dB higher, in agreement with the scaling law of $20\log N$ ($N=2$ or $3$). 
In comparison, we plot the results in Ref. \cite{Jin:21, Lihachev:22a, Liu:20} which use commercial, non-integrated PDs via fiber connection. 
Figure \ref{Fig:4}c shows the Allan deviations of the SIL laser frequency and the microwave frequency measured by the PNA. 
The resemblance in trend indicates that the long term stability of the microwave is dominated by the SIL laser frequency. 

%%%%%%%%%%%%%%%%%%%%%%%
\section*{Conclusion}
In conclusion, we have demonstrated a fully integrated photonic microwave oscillator at 10.7 GHz using a platicon microcomb. 
Such a hybrid chip device, occupying a footprint of only 76 mm$^2$, consists of a high-power DFB laser, a high-$Q$ Si$_3$N$_4$ microresonator, and a high-speed PD. 
Each component represents the state of the art of its own class, yet can be manufactured in large volume with low cost using established CMOS foundries of integrated photonics. 

Leveraging laser self-injection locking, the laser frequency noise is suppressed by 31 dB, yielding an intrinsic linewidth of 6.9 Hz and platicon microcomb generation. 
The PD chip further outputs a 10.7 GHz microwave carrier of 6.3 mHz linewidth, and harmonics of 10.7 GHz benefiting from the PD's 110 GHz bandwidth. 
In the future, the long-term stability of the three-in-one -- hybrid laser, comb and microwave -- can be improved by PDH-locking to an ultra-stable cavity \cite{Guo:22, Cheng:23}. 
Recent advances \cite{Yu:20, Xiang:21, Liu:20a, Snigirev:23} in heterogeneous integration of III-V, Si$_3$N$_4$ and modulators open a path towards chip devices of further reduced sizes, improved stability and added frequency tunability.
Up-shifting the microwave frequency, even into the millimeter-wave band, can be simply realized with Si$_3$N$_4$ microresonators of larger FSR \cite{Tetsumoto:21, Wang:21} and modified uni-traveling-carrier (MUTC) InP PD chips \cite{Li:23}. 
Such low-noise microwaves from hybrid photonic chips of small size, weight and power consumption can reinvigorate our information technology and applications for microwave photonics, terrestrial broadband, traffic control, tracking, analog-to-digital conversion, wireless networks, space links, and electron paramagnetic resonance spectroscopy.

\medskip
\begin{footnotesize}

\noindent\textbf{Methods}

\noindent \textbf{Characterization of microresonator dispersion.} 
We use a vector spectrum analyser (VSA) \cite{Luo:23} to measure the central frequency $\omega/2\pi$ and loaded linewidth $\kappa/2\pi=(\kappa_0 +\kappa_\text{ex})/2\pi$ of each resonance within 1480 to 1640 nm. 
The intrinsic loss $\kappa_0$ and external coupling rate $\kappa_\text{ex}$ can be fitted based on the resonance's transmission profile \cite{Gorodetsky:00, Li:13}, extinction ratio \cite{Cai:00} and phase response \cite{Luo:23}. 
The intrinsic ($Q_0$) and loaded ($Q_\text{m}$) quality factors of each resonance can be calculated as $Q_0=\omega/\kappa_0$ and $Q_\text{m}=\omega/\kappa$. 
Due to the presence of microresonator dispersion, the measured resonance grid deviates from an equidistant grid. 
Thus the $\mu^\text{th}$ resonance frequency $\omega_\mu$ can be written as 
\begin{align}
\omega_\mu=&\omega_0+D_1\mu+D_2\mu^2/2+D_3\mu^3/6+...\\
          =&\omega_0+D_1\mu+D_\text{int}
\end{align}
where $\omega_0$ is the reference resonance frequency, $D_1/2\pi$ is the microresonator FSR, $D_2$ describes the GVD, $D_3$ is higher-order dispersion terms. 
With known $\omega_\mu$, $\omega_0$ and fitted $D_1$, the microresonator's integrated dispersion $D_\text{int}=\omega_{\mu}-\omega_0-D_1\mu$ can be plotted. 

\vspace{0.1cm}
\noindent \textbf{Funding Information}: 
J. Liu acknowledges support from the National Natural Science Foundation of China (Grant No.12261131503), 
Shenzhen-Hong Kong Cooperation Zone for Technology and Innovation (HZQB-KCZYB2020050), 
and from the Guangdong Provincial Key Laboratory (2019B121203002). 
B. C. acknowledges support from the National Natural Science Foundation of China (Grant No. 61975121). 
Z. C. and Y.-H. L. acknowledge support from the China Postdoctoral Science Foundation (Grant No. 2023M731508 and 2022M721482). 

\vspace{0.1cm}
\noindent \textbf{Acknowledgments}: 
We thank Zhongkai Wang and Rumin Cheng for assistance in the experiment, Dengke Chen for assistance in data processing, Chengying Bao, Jijun He and Hairun Guo for the fruitful discussion. 
The DFB lasers were fabricated by Henan Shijia Photons Technology Co., Ltd and Shenzhen PhotonX Technology Co. Ltd. 
Silicon nitride chips were fabricated by Qaleido Photonics. 
The PD chips were fabricated with support from the ShanghaiTech University Quantum Device Lab (SQDL).

\vspace{0.1cm}
\noindent \textbf{Author contributions}: 
W. S., Z. C., J. Long, H. Z. performed the experiment with assistance from S. L.. 
The Si$_3$N$_4$ chips were fabricated by C. S. and L. G., and characterized by B. S. and Y.-H. L.. 
The PD chips were fabricated by L. L., L. W., Q. C. and B. C..
W. S., Z. C., Y.-H. L., B. C. and J. Liu wrote the manuscript, with the input from others.
J. Liu supervised the project.

\vspace{0.1cm}
\noindent \textbf{Disclosures}: 
W. S., Z. C., J. Long and J. Liu are inventors on a patent application related to this work. 
C. S. and J. Liu are co-founders of Qaleido Photonics, a start-up that is developing heterogeneous silicon nitride integrated photonics technologies. 
Others declare no conflicts of interest.

\vspace{0.1cm}
\noindent \textbf{Data Availability Statement}: 
The code and data used to produce the plots within this work will be released on the repository \texttt{Zenodo} upon publication of this preprint.

\end{footnotesize}
\bibliographystyle{apsrev4-1}
\bibliography{bibliography}

\begin{thebibliography}{64}%
\makeatletter
\providecommand \@ifxundefined [1]{%
 \@ifx{#1\undefined}
}%
\providecommand \@ifnum [1]{%
 \ifnum #1\expandafter \@firstoftwo
 \else \expandafter \@secondoftwo
 \fi
}%
\providecommand \@ifx [1]{%
 \ifx #1\expandafter \@firstoftwo
 \else \expandafter \@secondoftwo
 \fi
}%
\providecommand \natexlab [1]{#1}%
\providecommand \enquote  [1]{``#1''}%
\providecommand \bibnamefont  [1]{#1}%
\providecommand \bibfnamefont [1]{#1}%
\providecommand \citenamefont [1]{#1}%
\providecommand \href@noop [0]{\@secondoftwo}%
\providecommand \href [0]{\begingroup \@sanitize@url \@href}%
\providecommand \@href[1]{\@@startlink{#1}\@@href}%
\providecommand \@@href[1]{\endgroup#1\@@endlink}%
\providecommand \@sanitize@url [0]{\catcode `\\12\catcode `\$12\catcode `\&12\catcode `\#12\catcode `\^12\catcode `\_12\catcode `\%12\relax}%
\providecommand \@@startlink[1]{}%
\providecommand \@@endlink[0]{}%
\providecommand \url  [0]{\begingroup\@sanitize@url \@url }%
\providecommand \@url [1]{\endgroup\@href {#1}{\urlprefix }}%
\providecommand \urlprefix  [0]{URL }%
\providecommand \Eprint [0]{\href }%
\providecommand \doibase [0]{http://dx.doi.org/}%
\providecommand \selectlanguage [0]{\@gobble}%
\providecommand \bibinfo  [0]{\@secondoftwo}%
\providecommand \bibfield  [0]{\@secondoftwo}%
\providecommand \translation [1]{[#1]}%
\providecommand \BibitemOpen [0]{}%
\providecommand \bibitemStop [0]{}%
\providecommand \bibitemNoStop [0]{.\EOS\space}%
\providecommand \EOS [0]{\spacefactor3000\relax}%
\providecommand \BibitemShut  [1]{\csname bibitem#1\endcsname}%
\let\auto@bib@innerbib\@empty
%</preamble>
\bibitem [{\citenamefont {Fortier}\ \emph {et~al.}(2011)\citenamefont {Fortier}, \citenamefont {Kirchner}, \citenamefont {Quinlan}, \citenamefont {Taylor}, \citenamefont {Bergquist}, \citenamefont {Rosenband}, \citenamefont {Lemke}, \citenamefont {Ludlow}, \citenamefont {Jiang}, \citenamefont {Oates},\ and\ \citenamefont {Diddams}}]{Fortier:11}%
  \BibitemOpen
  \bibfield  {author} {\bibinfo {author} {\bibfnamefont {T.~M.}\ \bibnamefont {Fortier}}, \bibinfo {author} {\bibfnamefont {M.~S.}\ \bibnamefont {Kirchner}}, \bibinfo {author} {\bibfnamefont {F.}~\bibnamefont {Quinlan}}, \bibinfo {author} {\bibfnamefont {J.}~\bibnamefont {Taylor}}, \bibinfo {author} {\bibfnamefont {J.~C.}\ \bibnamefont {Bergquist}}, \bibinfo {author} {\bibfnamefont {T.}~\bibnamefont {Rosenband}}, \bibinfo {author} {\bibfnamefont {N.}~\bibnamefont {Lemke}}, \bibinfo {author} {\bibfnamefont {A.}~\bibnamefont {Ludlow}}, \bibinfo {author} {\bibfnamefont {Y.}~\bibnamefont {Jiang}}, \bibinfo {author} {\bibfnamefont {C.~W.}\ \bibnamefont {Oates}}, \ and\ \bibinfo {author} {\bibfnamefont {S.~A.}\ \bibnamefont {Diddams}},\ }\href {\doibase 10.1038/nphoton.2011.121} {\bibfield  {journal} {\bibinfo  {journal} {Nature Photonics}\ }\textbf {\bibinfo {volume} {5}},\ \bibinfo {pages} {425} (\bibinfo {year} {2011})}\BibitemShut {NoStop}%
\bibitem [{\citenamefont {Xie}\ \emph {et~al.}(2016)\citenamefont {Xie}, \citenamefont {Bouchand}, \citenamefont {Nicolodi}, \citenamefont {Giunta}, \citenamefont {H{\"a}nsel}, \citenamefont {Lezius}, \citenamefont {Joshi}, \citenamefont {Datta}, \citenamefont {Alexandre}, \citenamefont {Lours}, \citenamefont {Tremblin}, \citenamefont {Santarelli}, \citenamefont {Holzwarth},\ and\ \citenamefont {Le~Coq}}]{Xie:17}%
  \BibitemOpen
  \bibfield  {author} {\bibinfo {author} {\bibfnamefont {X.}~\bibnamefont {Xie}}, \bibinfo {author} {\bibfnamefont {R.}~\bibnamefont {Bouchand}}, \bibinfo {author} {\bibfnamefont {D.}~\bibnamefont {Nicolodi}}, \bibinfo {author} {\bibfnamefont {M.}~\bibnamefont {Giunta}}, \bibinfo {author} {\bibfnamefont {W.}~\bibnamefont {H{\"a}nsel}}, \bibinfo {author} {\bibfnamefont {M.}~\bibnamefont {Lezius}}, \bibinfo {author} {\bibfnamefont {A.}~\bibnamefont {Joshi}}, \bibinfo {author} {\bibfnamefont {S.}~\bibnamefont {Datta}}, \bibinfo {author} {\bibfnamefont {C.}~\bibnamefont {Alexandre}}, \bibinfo {author} {\bibfnamefont {M.}~\bibnamefont {Lours}}, \bibinfo {author} {\bibfnamefont {P.-A.}\ \bibnamefont {Tremblin}}, \bibinfo {author} {\bibfnamefont {G.}~\bibnamefont {Santarelli}}, \bibinfo {author} {\bibfnamefont {R.}~\bibnamefont {Holzwarth}}, \ and\ \bibinfo {author} {\bibfnamefont {Y.}~\bibnamefont {Le~Coq}},\ }\href {https://doi.org/10.1038/nphoton.2016.215} {\bibfield  {journal} {\bibinfo  {journal} {Nature
  Photonics}\ }\textbf {\bibinfo {volume} {11}},\ \bibinfo {pages} {44} (\bibinfo {year} {2016})}\BibitemShut {NoStop}%
\bibitem [{\citenamefont {Li}\ \emph {et~al.}(2014)\citenamefont {Li}, \citenamefont {Yi}, \citenamefont {Lee}, \citenamefont {Diddams},\ and\ \citenamefont {Vahala}}]{Li:14}%
  \BibitemOpen
  \bibfield  {author} {\bibinfo {author} {\bibfnamefont {J.}~\bibnamefont {Li}}, \bibinfo {author} {\bibfnamefont {X.}~\bibnamefont {Yi}}, \bibinfo {author} {\bibfnamefont {H.}~\bibnamefont {Lee}}, \bibinfo {author} {\bibfnamefont {S.~A.}\ \bibnamefont {Diddams}}, \ and\ \bibinfo {author} {\bibfnamefont {K.~J.}\ \bibnamefont {Vahala}},\ }\href {\doibase 10.1126/science.1252909} {\bibfield  {journal} {\bibinfo  {journal} {Science}\ }\textbf {\bibinfo {volume} {345}},\ \bibinfo {pages} {309} (\bibinfo {year} {2014})}\BibitemShut {NoStop}%
\bibitem [{\citenamefont {Nakamura}\ \emph {et~al.}(2020)\citenamefont {Nakamura}, \citenamefont {Davila-Rodriguez}, \citenamefont {Leopardi}, \citenamefont {Sherman}, \citenamefont {Fortier}, \citenamefont {Xie}, \citenamefont {Campbell}, \citenamefont {McGrew}, \citenamefont {Zhang}, \citenamefont {Hassan}, \citenamefont {Nicolodi}, \citenamefont {Beloy}, \citenamefont {Ludlow}, \citenamefont {Diddams},\ and\ \citenamefont {Quinlan}}]{Nakamura:20}%
  \BibitemOpen
  \bibfield  {author} {\bibinfo {author} {\bibfnamefont {T.}~\bibnamefont {Nakamura}}, \bibinfo {author} {\bibfnamefont {J.}~\bibnamefont {Davila-Rodriguez}}, \bibinfo {author} {\bibfnamefont {H.}~\bibnamefont {Leopardi}}, \bibinfo {author} {\bibfnamefont {J.~A.}\ \bibnamefont {Sherman}}, \bibinfo {author} {\bibfnamefont {T.~M.}\ \bibnamefont {Fortier}}, \bibinfo {author} {\bibfnamefont {X.}~\bibnamefont {Xie}}, \bibinfo {author} {\bibfnamefont {J.~C.}\ \bibnamefont {Campbell}}, \bibinfo {author} {\bibfnamefont {W.~F.}\ \bibnamefont {McGrew}}, \bibinfo {author} {\bibfnamefont {X.}~\bibnamefont {Zhang}}, \bibinfo {author} {\bibfnamefont {Y.~S.}\ \bibnamefont {Hassan}}, \bibinfo {author} {\bibfnamefont {D.}~\bibnamefont {Nicolodi}}, \bibinfo {author} {\bibfnamefont {K.}~\bibnamefont {Beloy}}, \bibinfo {author} {\bibfnamefont {A.~D.}\ \bibnamefont {Ludlow}}, \bibinfo {author} {\bibfnamefont {S.~A.}\ \bibnamefont {Diddams}}, \ and\ \bibinfo {author} {\bibfnamefont {F.}~\bibnamefont {Quinlan}},\ }\href {\doibase
  10.1126/science.abb2473} {\bibfield  {journal} {\bibinfo  {journal} {Science}\ }\textbf {\bibinfo {volume} {368}},\ \bibinfo {pages} {889} (\bibinfo {year} {2020})}\BibitemShut {NoStop}%
\bibitem [{\citenamefont {Kippenberg}\ \emph {et~al.}(2018)\citenamefont {Kippenberg}, \citenamefont {Gaeta}, \citenamefont {Lipson},\ and\ \citenamefont {Gorodetsky}}]{Kippenberg:18}%
  \BibitemOpen
  \bibfield  {author} {\bibinfo {author} {\bibfnamefont {T.~J.}\ \bibnamefont {Kippenberg}}, \bibinfo {author} {\bibfnamefont {A.~L.}\ \bibnamefont {Gaeta}}, \bibinfo {author} {\bibfnamefont {M.}~\bibnamefont {Lipson}}, \ and\ \bibinfo {author} {\bibfnamefont {M.~L.}\ \bibnamefont {Gorodetsky}},\ }\href {\doibase 10.1126/science.aan8083} {\bibfield  {journal} {\bibinfo  {journal} {Science}\ }\textbf {\bibinfo {volume} {361}},\ \bibinfo {pages} {eaan8083} (\bibinfo {year} {2018})}\BibitemShut {NoStop}%
\bibitem [{\citenamefont {Pasquazi}\ \emph {et~al.}(2018)\citenamefont {Pasquazi}, \citenamefont {Peccianti}, \citenamefont {Razzari}, \citenamefont {Moss}, \citenamefont {Coen}, \citenamefont {Erkintalo}, \citenamefont {Chembo}, \citenamefont {Hansson}, \citenamefont {Wabnitz}, \citenamefont {Del'Haye}, \citenamefont {Xue}, \citenamefont {Weiner},\ and\ \citenamefont {Morandotti}}]{Pasquazi:18}%
  \BibitemOpen
  \bibfield  {author} {\bibinfo {author} {\bibfnamefont {A.}~\bibnamefont {Pasquazi}}, \bibinfo {author} {\bibfnamefont {M.}~\bibnamefont {Peccianti}}, \bibinfo {author} {\bibfnamefont {L.}~\bibnamefont {Razzari}}, \bibinfo {author} {\bibfnamefont {D.~J.}\ \bibnamefont {Moss}}, \bibinfo {author} {\bibfnamefont {S.}~\bibnamefont {Coen}}, \bibinfo {author} {\bibfnamefont {M.}~\bibnamefont {Erkintalo}}, \bibinfo {author} {\bibfnamefont {Y.~K.}\ \bibnamefont {Chembo}}, \bibinfo {author} {\bibfnamefont {T.}~\bibnamefont {Hansson}}, \bibinfo {author} {\bibfnamefont {S.}~\bibnamefont {Wabnitz}}, \bibinfo {author} {\bibfnamefont {P.}~\bibnamefont {Del'Haye}}, \bibinfo {author} {\bibfnamefont {X.}~\bibnamefont {Xue}}, \bibinfo {author} {\bibfnamefont {A.~M.}\ \bibnamefont {Weiner}}, \ and\ \bibinfo {author} {\bibfnamefont {R.}~\bibnamefont {Morandotti}},\ }\href {\doibase https://doi.org/10.1016/j.physrep.2017.08.004} {\bibfield  {journal} {\bibinfo  {journal} {Physics Reports}\ }\textbf {\bibinfo {volume} {729}},\
  \bibinfo {pages} {1} (\bibinfo {year} {2018})}\BibitemShut {NoStop}%
\bibitem [{\citenamefont {Moss}\ \emph {et~al.}(2013)\citenamefont {Moss}, \citenamefont {Morandotti}, \citenamefont {Gaeta},\ and\ \citenamefont {Lipson}}]{Moss:13}%
  \BibitemOpen
  \bibfield  {author} {\bibinfo {author} {\bibfnamefont {D.~J.}\ \bibnamefont {Moss}}, \bibinfo {author} {\bibfnamefont {R.}~\bibnamefont {Morandotti}}, \bibinfo {author} {\bibfnamefont {A.~L.}\ \bibnamefont {Gaeta}}, \ and\ \bibinfo {author} {\bibfnamefont {M.}~\bibnamefont {Lipson}},\ }\href {https://doi.org/10.1038/nphoton.2013.183} {\bibfield  {journal} {\bibinfo  {journal} {Nature Photonics}\ }\textbf {\bibinfo {volume} {7}},\ \bibinfo {pages} {597} (\bibinfo {year} {2013})}\BibitemShut {NoStop}%
\bibitem [{\citenamefont {Gaeta}\ \emph {et~al.}(2019)\citenamefont {Gaeta}, \citenamefont {Lipson},\ and\ \citenamefont {Kippenberg}}]{Gaeta:19}%
  \BibitemOpen
  \bibfield  {author} {\bibinfo {author} {\bibfnamefont {A.~L.}\ \bibnamefont {Gaeta}}, \bibinfo {author} {\bibfnamefont {M.}~\bibnamefont {Lipson}}, \ and\ \bibinfo {author} {\bibfnamefont {T.~J.}\ \bibnamefont {Kippenberg}},\ }\href {\doibase 10.1038/s41566-019-0358-x} {\bibfield  {journal} {\bibinfo  {journal} {Nature Photonics}\ }\textbf {\bibinfo {volume} {13}},\ \bibinfo {pages} {158} (\bibinfo {year} {2019})}\BibitemShut {NoStop}%
\bibitem [{\citenamefont {Liu}\ \emph {et~al.}(2021)\citenamefont {Liu}, \citenamefont {Huang}, \citenamefont {Wang}, \citenamefont {He}, \citenamefont {Raja}, \citenamefont {Liu}, \citenamefont {Engelsen},\ and\ \citenamefont {Kippenberg}}]{Liu:21}%
  \BibitemOpen
  \bibfield  {author} {\bibinfo {author} {\bibfnamefont {J.}~\bibnamefont {Liu}}, \bibinfo {author} {\bibfnamefont {G.}~\bibnamefont {Huang}}, \bibinfo {author} {\bibfnamefont {R.~N.}\ \bibnamefont {Wang}}, \bibinfo {author} {\bibfnamefont {J.}~\bibnamefont {He}}, \bibinfo {author} {\bibfnamefont {A.~S.}\ \bibnamefont {Raja}}, \bibinfo {author} {\bibfnamefont {T.}~\bibnamefont {Liu}}, \bibinfo {author} {\bibfnamefont {N.~J.}\ \bibnamefont {Engelsen}}, \ and\ \bibinfo {author} {\bibfnamefont {T.~J.}\ \bibnamefont {Kippenberg}},\ }\href {\doibase 10.1038/s41467-021-21973-z} {\bibfield  {journal} {\bibinfo  {journal} {Nature Communications}\ }\textbf {\bibinfo {volume} {12}},\ \bibinfo {pages} {2236} (\bibinfo {year} {2021})}\BibitemShut {NoStop}%
\bibitem [{\citenamefont {Liang}\ \emph {et~al.}(2015{\natexlab{a}})\citenamefont {Liang}, \citenamefont {Eliyahu}, \citenamefont {Ilchenko}, \citenamefont {Savchenkov}, \citenamefont {Matsko}, \citenamefont {Seidel},\ and\ \citenamefont {Maleki}}]{Liang:15}%
  \BibitemOpen
  \bibfield  {author} {\bibinfo {author} {\bibfnamefont {W.}~\bibnamefont {Liang}}, \bibinfo {author} {\bibfnamefont {D.}~\bibnamefont {Eliyahu}}, \bibinfo {author} {\bibfnamefont {V.~S.}\ \bibnamefont {Ilchenko}}, \bibinfo {author} {\bibfnamefont {A.~A.}\ \bibnamefont {Savchenkov}}, \bibinfo {author} {\bibfnamefont {A.~B.}\ \bibnamefont {Matsko}}, \bibinfo {author} {\bibfnamefont {D.}~\bibnamefont {Seidel}}, \ and\ \bibinfo {author} {\bibfnamefont {L.}~\bibnamefont {Maleki}},\ }\href {https://doi.org/10.1038/ncomms8957} {\bibfield  {journal} {\bibinfo  {journal} {Nature Communications}\ }\textbf {\bibinfo {volume} {6}},\ \bibinfo {pages} {7957} (\bibinfo {year} {2015}{\natexlab{a}})}\BibitemShut {NoStop}%
\bibitem [{\citenamefont {Liu}\ \emph {et~al.}(2020{\natexlab{a}})\citenamefont {Liu}, \citenamefont {Lucas}, \citenamefont {Raja}, \citenamefont {He}, \citenamefont {Riemensberger}, \citenamefont {Wang}, \citenamefont {Karpov}, \citenamefont {Guo}, \citenamefont {Bouchand},\ and\ \citenamefont {Kippenberg}}]{Liu:20}%
  \BibitemOpen
  \bibfield  {author} {\bibinfo {author} {\bibfnamefont {J.}~\bibnamefont {Liu}}, \bibinfo {author} {\bibfnamefont {E.}~\bibnamefont {Lucas}}, \bibinfo {author} {\bibfnamefont {A.~S.}\ \bibnamefont {Raja}}, \bibinfo {author} {\bibfnamefont {J.}~\bibnamefont {He}}, \bibinfo {author} {\bibfnamefont {J.}~\bibnamefont {Riemensberger}}, \bibinfo {author} {\bibfnamefont {R.~N.}\ \bibnamefont {Wang}}, \bibinfo {author} {\bibfnamefont {M.}~\bibnamefont {Karpov}}, \bibinfo {author} {\bibfnamefont {H.}~\bibnamefont {Guo}}, \bibinfo {author} {\bibfnamefont {R.}~\bibnamefont {Bouchand}}, \ and\ \bibinfo {author} {\bibfnamefont {T.~J.}\ \bibnamefont {Kippenberg}},\ }\href {\doibase 10.1038/s41566-020-0617-x} {\bibfield  {journal} {\bibinfo  {journal} {Nature Photonics}\ }\textbf {\bibinfo {volume} {14}},\ \bibinfo {pages} {486} (\bibinfo {year} {2020}{\natexlab{a}})}\BibitemShut {NoStop}%
\bibitem [{\citenamefont {Yao}\ \emph {et~al.}(2022)\citenamefont {Yao}, \citenamefont {Liu}, \citenamefont {Chen}, \citenamefont {Gong}, \citenamefont {Yang},\ and\ \citenamefont {Xiao}}]{Yao:22}%
  \BibitemOpen
  \bibfield  {author} {\bibinfo {author} {\bibfnamefont {L.}~\bibnamefont {Yao}}, \bibinfo {author} {\bibfnamefont {P.}~\bibnamefont {Liu}}, \bibinfo {author} {\bibfnamefont {H.-J.}\ \bibnamefont {Chen}}, \bibinfo {author} {\bibfnamefont {Q.}~\bibnamefont {Gong}}, \bibinfo {author} {\bibfnamefont {Q.-F.}\ \bibnamefont {Yang}}, \ and\ \bibinfo {author} {\bibfnamefont {Y.-F.}\ \bibnamefont {Xiao}},\ }\href {\doibase 10.1364/OPTICA.459130} {\bibfield  {journal} {\bibinfo  {journal} {Optica}\ }\textbf {\bibinfo {volume} {9}},\ \bibinfo {pages} {561} (\bibinfo {year} {2022})}\BibitemShut {NoStop}%
\bibitem [{\citenamefont {Tetsumoto}\ \emph {et~al.}(2021)\citenamefont {Tetsumoto}, \citenamefont {Nagatsuma}, \citenamefont {Fermann}, \citenamefont {Navickaite}, \citenamefont {Geiselmann},\ and\ \citenamefont {Rolland}}]{Tetsumoto:21}%
  \BibitemOpen
  \bibfield  {author} {\bibinfo {author} {\bibfnamefont {T.}~\bibnamefont {Tetsumoto}}, \bibinfo {author} {\bibfnamefont {T.}~\bibnamefont {Nagatsuma}}, \bibinfo {author} {\bibfnamefont {M.~E.}\ \bibnamefont {Fermann}}, \bibinfo {author} {\bibfnamefont {G.}~\bibnamefont {Navickaite}}, \bibinfo {author} {\bibfnamefont {M.}~\bibnamefont {Geiselmann}}, \ and\ \bibinfo {author} {\bibfnamefont {A.}~\bibnamefont {Rolland}},\ }\href {\doibase 10.1038/s41566-021-00790-2} {\bibfield  {journal} {\bibinfo  {journal} {Nature Photonics}\ }\textbf {\bibinfo {volume} {15}},\ \bibinfo {pages} {516} (\bibinfo {year} {2021})}\BibitemShut {NoStop}%
\bibitem [{\citenamefont {Wang}\ \emph {et~al.}(2021)\citenamefont {Wang}, \citenamefont {Morgan}, \citenamefont {Sun}, \citenamefont {Jahanbozorgi}, \citenamefont {Yang}, \citenamefont {Woodson}, \citenamefont {Estrella}, \citenamefont {Beling},\ and\ \citenamefont {Yi}}]{Wang:21}%
  \BibitemOpen
  \bibfield  {author} {\bibinfo {author} {\bibfnamefont {B.}~\bibnamefont {Wang}}, \bibinfo {author} {\bibfnamefont {J.~S.}\ \bibnamefont {Morgan}}, \bibinfo {author} {\bibfnamefont {K.}~\bibnamefont {Sun}}, \bibinfo {author} {\bibfnamefont {M.}~\bibnamefont {Jahanbozorgi}}, \bibinfo {author} {\bibfnamefont {Z.}~\bibnamefont {Yang}}, \bibinfo {author} {\bibfnamefont {M.}~\bibnamefont {Woodson}}, \bibinfo {author} {\bibfnamefont {S.}~\bibnamefont {Estrella}}, \bibinfo {author} {\bibfnamefont {A.}~\bibnamefont {Beling}}, \ and\ \bibinfo {author} {\bibfnamefont {X.}~\bibnamefont {Yi}},\ }\href {\doibase 10.1038/s41377-020-00445-x} {\bibfield  {journal} {\bibinfo  {journal} {Light: Science \& Applications}\ }\textbf {\bibinfo {volume} {10}},\ \bibinfo {pages} {4} (\bibinfo {year} {2021})}\BibitemShut {NoStop}%
\bibitem [{\citenamefont {Li}\ \emph {et~al.}(2013{\natexlab{a}})\citenamefont {Li}, \citenamefont {Lee},\ and\ \citenamefont {Vahala}}]{LiJ:13}%
  \BibitemOpen
  \bibfield  {author} {\bibinfo {author} {\bibfnamefont {J.}~\bibnamefont {Li}}, \bibinfo {author} {\bibfnamefont {H.}~\bibnamefont {Lee}}, \ and\ \bibinfo {author} {\bibfnamefont {K.~J.}\ \bibnamefont {Vahala}},\ }\href {https://doi.org/10.1038/ncomms3097} {\bibfield  {journal} {\bibinfo  {journal} {Nature Communications}\ }\textbf {\bibinfo {volume} {4}},\ \bibinfo {pages} {2097} (\bibinfo {year} {2013}{\natexlab{a}})}\BibitemShut {NoStop}%
\bibitem [{\citenamefont {Li}\ and\ \citenamefont {Vahala}(2023)}]{LiJ:23}%
  \BibitemOpen
  \bibfield  {author} {\bibinfo {author} {\bibfnamefont {J.}~\bibnamefont {Li}}\ and\ \bibinfo {author} {\bibfnamefont {K.}~\bibnamefont {Vahala}},\ }\href {\doibase 10.1364/OPTICA.477602} {\bibfield  {journal} {\bibinfo  {journal} {Optica}\ }\textbf {\bibinfo {volume} {10}},\ \bibinfo {pages} {33} (\bibinfo {year} {2023})}\BibitemShut {NoStop}%
\bibitem [{\citenamefont {Tang}\ \emph {et~al.}(2018)\citenamefont {Tang}, \citenamefont {Hao}, \citenamefont {Li}, \citenamefont {Domenech}, \citenamefont {{n}os}, \citenamefont {{n}oz}, \citenamefont {Zhu}, \citenamefont {Capmany},\ and\ \citenamefont {Li}}]{Tang:18}%
  \BibitemOpen
  \bibfield  {author} {\bibinfo {author} {\bibfnamefont {J.}~\bibnamefont {Tang}}, \bibinfo {author} {\bibfnamefont {T.}~\bibnamefont {Hao}}, \bibinfo {author} {\bibfnamefont {W.}~\bibnamefont {Li}}, \bibinfo {author} {\bibfnamefont {D.}~\bibnamefont {Domenech}}, \bibinfo {author} {\bibfnamefont {R.~B.}\ \bibnamefont {{n}os}}, \bibinfo {author} {\bibfnamefont {P.~M.}\ \bibnamefont {{n}oz}}, \bibinfo {author} {\bibfnamefont {N.}~\bibnamefont {Zhu}}, \bibinfo {author} {\bibfnamefont {J.}~\bibnamefont {Capmany}}, \ and\ \bibinfo {author} {\bibfnamefont {M.}~\bibnamefont {Li}},\ }\href {\doibase 10.1364/OE.26.012257} {\bibfield  {journal} {\bibinfo  {journal} {Opt. Express}\ }\textbf {\bibinfo {volume} {26}},\ \bibinfo {pages} {12257} (\bibinfo {year} {2018})}\BibitemShut {NoStop}%
\bibitem [{\citenamefont {Cundiff}\ and\ \citenamefont {Ye}(2003)}]{Cundiff:03}%
  \BibitemOpen
  \bibfield  {author} {\bibinfo {author} {\bibfnamefont {S.~T.}\ \bibnamefont {Cundiff}}\ and\ \bibinfo {author} {\bibfnamefont {J.}~\bibnamefont {Ye}},\ }\href {\doibase 10.1103/RevModPhys.75.325} {\bibfield  {journal} {\bibinfo  {journal} {Rev. Mod. Phys.}\ }\textbf {\bibinfo {volume} {75}},\ \bibinfo {pages} {325} (\bibinfo {year} {2003})}\BibitemShut {NoStop}%
\bibitem [{\citenamefont {Fortier}\ and\ \citenamefont {Baumann}(2019)}]{Fortier:19}%
  \BibitemOpen
  \bibfield  {author} {\bibinfo {author} {\bibfnamefont {T.}~\bibnamefont {Fortier}}\ and\ \bibinfo {author} {\bibfnamefont {E.}~\bibnamefont {Baumann}},\ }\href {\doibase 10.1038/s42005-019-0249-y} {\bibfield  {journal} {\bibinfo  {journal} {Communications Physics}\ }\textbf {\bibinfo {volume} {2}},\ \bibinfo {pages} {153} (\bibinfo {year} {2019})}\BibitemShut {NoStop}%
\bibitem [{\citenamefont {Diddams}\ \emph {et~al.}(2020)\citenamefont {Diddams}, \citenamefont {Vahala},\ and\ \citenamefont {Udem}}]{Diddams:20}%
  \BibitemOpen
  \bibfield  {author} {\bibinfo {author} {\bibfnamefont {S.~A.}\ \bibnamefont {Diddams}}, \bibinfo {author} {\bibfnamefont {K.}~\bibnamefont {Vahala}}, \ and\ \bibinfo {author} {\bibfnamefont {T.}~\bibnamefont {Udem}},\ }\href {\doibase 10.1126/science.aay3676} {\bibfield  {journal} {\bibinfo  {journal} {Science}\ }\textbf {\bibinfo {volume} {369}},\ \bibinfo {pages} {eaay3676} (\bibinfo {year} {2020})}\BibitemShut {NoStop}%
\bibitem [{\citenamefont {Herr}\ \emph {et~al.}(2013)\citenamefont {Herr}, \citenamefont {Brasch}, \citenamefont {Jost}, \citenamefont {Wang}, \citenamefont {Kondratiev}, \citenamefont {Gorodetsky},\ and\ \citenamefont {Kippenberg}}]{Herr:14}%
  \BibitemOpen
  \bibfield  {author} {\bibinfo {author} {\bibfnamefont {T.}~\bibnamefont {Herr}}, \bibinfo {author} {\bibfnamefont {V.}~\bibnamefont {Brasch}}, \bibinfo {author} {\bibfnamefont {J.~D.}\ \bibnamefont {Jost}}, \bibinfo {author} {\bibfnamefont {C.~Y.}\ \bibnamefont {Wang}}, \bibinfo {author} {\bibfnamefont {N.~M.}\ \bibnamefont {Kondratiev}}, \bibinfo {author} {\bibfnamefont {M.~L.}\ \bibnamefont {Gorodetsky}}, \ and\ \bibinfo {author} {\bibfnamefont {T.~J.}\ \bibnamefont {Kippenberg}},\ }\href {https://doi.org/10.1038/nphoton.2013.343} {\bibfield  {journal} {\bibinfo  {journal} {Nature Photonics}\ }\textbf {\bibinfo {volume} {8}},\ \bibinfo {pages} {145} (\bibinfo {year} {2013})}\BibitemShut {NoStop}%
\bibitem [{\citenamefont {Yi}\ \emph {et~al.}(2015)\citenamefont {Yi}, \citenamefont {Yang}, \citenamefont {Yang}, \citenamefont {Suh},\ and\ \citenamefont {Vahala}}]{Yi:15}%
  \BibitemOpen
  \bibfield  {author} {\bibinfo {author} {\bibfnamefont {X.}~\bibnamefont {Yi}}, \bibinfo {author} {\bibfnamefont {Q.-F.}\ \bibnamefont {Yang}}, \bibinfo {author} {\bibfnamefont {K.~Y.}\ \bibnamefont {Yang}}, \bibinfo {author} {\bibfnamefont {M.-G.}\ \bibnamefont {Suh}}, \ and\ \bibinfo {author} {\bibfnamefont {K.}~\bibnamefont {Vahala}},\ }\href {\doibase 10.1364/OPTICA.2.001078} {\bibfield  {journal} {\bibinfo  {journal} {Optica}\ }\textbf {\bibinfo {volume} {2}},\ \bibinfo {pages} {1078} (\bibinfo {year} {2015})}\BibitemShut {NoStop}%
\bibitem [{\citenamefont {Brasch}\ \emph {et~al.}(2016)\citenamefont {Brasch}, \citenamefont {Geiselmann}, \citenamefont {Herr}, \citenamefont {Lihachev}, \citenamefont {Pfeiffer}, \citenamefont {Gorodetsky},\ and\ \citenamefont {Kippenberg}}]{Brasch:15}%
  \BibitemOpen
  \bibfield  {author} {\bibinfo {author} {\bibfnamefont {V.}~\bibnamefont {Brasch}}, \bibinfo {author} {\bibfnamefont {M.}~\bibnamefont {Geiselmann}}, \bibinfo {author} {\bibfnamefont {T.}~\bibnamefont {Herr}}, \bibinfo {author} {\bibfnamefont {G.}~\bibnamefont {Lihachev}}, \bibinfo {author} {\bibfnamefont {M.~H.~P.}\ \bibnamefont {Pfeiffer}}, \bibinfo {author} {\bibfnamefont {M.~L.}\ \bibnamefont {Gorodetsky}}, \ and\ \bibinfo {author} {\bibfnamefont {T.~J.}\ \bibnamefont {Kippenberg}},\ }\href {\doibase 10.1126/science.aad4811} {\bibfield  {journal} {\bibinfo  {journal} {Science}\ }\textbf {\bibinfo {volume} {351}},\ \bibinfo {pages} {357} (\bibinfo {year} {2016})}\BibitemShut {NoStop}%
\bibitem [{\citenamefont {Joshi}\ \emph {et~al.}(2016)\citenamefont {Joshi}, \citenamefont {Jang}, \citenamefont {Luke}, \citenamefont {Ji}, \citenamefont {Miller}, \citenamefont {Klenner}, \citenamefont {Okawachi}, \citenamefont {Lipson},\ and\ \citenamefont {Gaeta}}]{Joshi:16}%
  \BibitemOpen
  \bibfield  {author} {\bibinfo {author} {\bibfnamefont {C.}~\bibnamefont {Joshi}}, \bibinfo {author} {\bibfnamefont {J.~K.}\ \bibnamefont {Jang}}, \bibinfo {author} {\bibfnamefont {K.}~\bibnamefont {Luke}}, \bibinfo {author} {\bibfnamefont {X.}~\bibnamefont {Ji}}, \bibinfo {author} {\bibfnamefont {S.~A.}\ \bibnamefont {Miller}}, \bibinfo {author} {\bibfnamefont {A.}~\bibnamefont {Klenner}}, \bibinfo {author} {\bibfnamefont {Y.}~\bibnamefont {Okawachi}}, \bibinfo {author} {\bibfnamefont {M.}~\bibnamefont {Lipson}}, \ and\ \bibinfo {author} {\bibfnamefont {A.~L.}\ \bibnamefont {Gaeta}},\ }\href {\doibase 10.1364/OL.41.002565} {\bibfield  {journal} {\bibinfo  {journal} {Opt. Lett.}\ }\textbf {\bibinfo {volume} {41}},\ \bibinfo {pages} {2565} (\bibinfo {year} {2016})}\BibitemShut {NoStop}%
\bibitem [{\citenamefont {Xue}\ \emph {et~al.}(2015)\citenamefont {Xue}, \citenamefont {Xuan}, \citenamefont {Liu}, \citenamefont {Wang}, \citenamefont {Chen}, \citenamefont {Wang}, \citenamefont {Leaird}, \citenamefont {Qi},\ and\ \citenamefont {Weiner}}]{Xue:15}%
  \BibitemOpen
  \bibfield  {author} {\bibinfo {author} {\bibfnamefont {X.}~\bibnamefont {Xue}}, \bibinfo {author} {\bibfnamefont {Y.}~\bibnamefont {Xuan}}, \bibinfo {author} {\bibfnamefont {Y.}~\bibnamefont {Liu}}, \bibinfo {author} {\bibfnamefont {P.-H.}\ \bibnamefont {Wang}}, \bibinfo {author} {\bibfnamefont {S.}~\bibnamefont {Chen}}, \bibinfo {author} {\bibfnamefont {J.}~\bibnamefont {Wang}}, \bibinfo {author} {\bibfnamefont {D.~E.}\ \bibnamefont {Leaird}}, \bibinfo {author} {\bibfnamefont {M.}~\bibnamefont {Qi}}, \ and\ \bibinfo {author} {\bibfnamefont {A.~M.}\ \bibnamefont {Weiner}},\ }\href {http://dx.doi.org/10.1038/nphoton.2015.137} {\bibfield  {journal} {\bibinfo  {journal} {Nature Photonics}\ }\textbf {\bibinfo {volume} {9}},\ \bibinfo {pages} {594} (\bibinfo {year} {2015})}\BibitemShut {NoStop}%
\bibitem [{\citenamefont {Kovach}\ \emph {et~al.}(2020)\citenamefont {Kovach}, \citenamefont {Chen}, \citenamefont {He}, \citenamefont {Choi}, \citenamefont {Dogan}, \citenamefont {Ghasemkhani}, \citenamefont {Taheri},\ and\ \citenamefont {Armani}}]{Kovach:20}%
  \BibitemOpen
  \bibfield  {author} {\bibinfo {author} {\bibfnamefont {A.}~\bibnamefont {Kovach}}, \bibinfo {author} {\bibfnamefont {D.}~\bibnamefont {Chen}}, \bibinfo {author} {\bibfnamefont {J.}~\bibnamefont {He}}, \bibinfo {author} {\bibfnamefont {H.}~\bibnamefont {Choi}}, \bibinfo {author} {\bibfnamefont {A.~H.}\ \bibnamefont {Dogan}}, \bibinfo {author} {\bibfnamefont {M.}~\bibnamefont {Ghasemkhani}}, \bibinfo {author} {\bibfnamefont {H.}~\bibnamefont {Taheri}}, \ and\ \bibinfo {author} {\bibfnamefont {A.~M.}\ \bibnamefont {Armani}},\ }\href {\doibase 10.1364/AOP.376924} {\bibfield  {journal} {\bibinfo  {journal} {Adv. Opt. Photon.}\ }\textbf {\bibinfo {volume} {12}},\ \bibinfo {pages} {135} (\bibinfo {year} {2020})}\BibitemShut {NoStop}%
\bibitem [{\citenamefont {Chang}\ \emph {et~al.}(2022)\citenamefont {Chang}, \citenamefont {Liu},\ and\ \citenamefont {Bowers}}]{Chang:22}%
  \BibitemOpen
  \bibfield  {author} {\bibinfo {author} {\bibfnamefont {L.}~\bibnamefont {Chang}}, \bibinfo {author} {\bibfnamefont {S.}~\bibnamefont {Liu}}, \ and\ \bibinfo {author} {\bibfnamefont {J.~E.}\ \bibnamefont {Bowers}},\ }\href {\doibase 10.1038/s41566-021-00945-1} {\bibfield  {journal} {\bibinfo  {journal} {Nature Photonics}\ }\textbf {\bibinfo {volume} {16}},\ \bibinfo {pages} {95} (\bibinfo {year} {2022})}\BibitemShut {NoStop}%
\bibitem [{\citenamefont {Kudelin}\ \emph {et~al.}(2023)\citenamefont {Kudelin}, \citenamefont {Groman}, \citenamefont {Ji}, \citenamefont {Guo}, \citenamefont {Kelleher}, \citenamefont {Lee}, \citenamefont {Nakamura}, \citenamefont {McLemore}, \citenamefont {Shirmohammadi}, \citenamefont {Hanifi}, \citenamefont {Cheng}, \citenamefont {Jin}, \citenamefont {Halliday}, \citenamefont {Dai}, \citenamefont {Wu}, \citenamefont {Jin}, \citenamefont {Liu}, \citenamefont {Zhang}, \citenamefont {Xiang}, \citenamefont {Iltchenko}, \citenamefont {Miller}, \citenamefont {Matsko}, \citenamefont {Bowers}, \citenamefont {Rakich}, \citenamefont {Campbell}, \citenamefont {Bowers}, \citenamefont {Vahala}, \citenamefont {Quinlan},\ and\ \citenamefont {Diddams}}]{Kudelin:23}%
  \BibitemOpen
  \bibfield  {author} {\bibinfo {author} {\bibfnamefont {I.}~\bibnamefont {Kudelin}}, \bibinfo {author} {\bibfnamefont {W.}~\bibnamefont {Groman}}, \bibinfo {author} {\bibfnamefont {Q.-X.}\ \bibnamefont {Ji}}, \bibinfo {author} {\bibfnamefont {J.}~\bibnamefont {Guo}}, \bibinfo {author} {\bibfnamefont {M.~L.}\ \bibnamefont {Kelleher}}, \bibinfo {author} {\bibfnamefont {D.}~\bibnamefont {Lee}}, \bibinfo {author} {\bibfnamefont {T.}~\bibnamefont {Nakamura}}, \bibinfo {author} {\bibfnamefont {C.~A.}\ \bibnamefont {McLemore}}, \bibinfo {author} {\bibfnamefont {P.}~\bibnamefont {Shirmohammadi}}, \bibinfo {author} {\bibfnamefont {S.}~\bibnamefont {Hanifi}}, \bibinfo {author} {\bibfnamefont {H.}~\bibnamefont {Cheng}}, \bibinfo {author} {\bibfnamefont {N.}~\bibnamefont {Jin}}, \bibinfo {author} {\bibfnamefont {S.}~\bibnamefont {Halliday}}, \bibinfo {author} {\bibfnamefont {Z.}~\bibnamefont {Dai}}, \bibinfo {author} {\bibfnamefont {L.}~\bibnamefont {Wu}}, \bibinfo {author} {\bibfnamefont {W.}~\bibnamefont {Jin}},
  \bibinfo {author} {\bibfnamefont {Y.}~\bibnamefont {Liu}}, \bibinfo {author} {\bibfnamefont {W.}~\bibnamefont {Zhang}}, \bibinfo {author} {\bibfnamefont {C.}~\bibnamefont {Xiang}}, \bibinfo {author} {\bibfnamefont {V.}~\bibnamefont {Iltchenko}}, \bibinfo {author} {\bibfnamefont {O.}~\bibnamefont {Miller}}, \bibinfo {author} {\bibfnamefont {A.}~\bibnamefont {Matsko}}, \bibinfo {author} {\bibfnamefont {S.}~\bibnamefont {Bowers}}, \bibinfo {author} {\bibfnamefont {P.~T.}\ \bibnamefont {Rakich}}, \bibinfo {author} {\bibfnamefont {J.~C.}\ \bibnamefont {Campbell}}, \bibinfo {author} {\bibfnamefont {J.~E.}\ \bibnamefont {Bowers}}, \bibinfo {author} {\bibfnamefont {K.}~\bibnamefont {Vahala}}, \bibinfo {author} {\bibfnamefont {F.}~\bibnamefont {Quinlan}}, \ and\ \bibinfo {author} {\bibfnamefont {S.~A.}\ \bibnamefont {Diddams}},\ }\href@noop {} {\bibfield  {journal} {\bibinfo  {journal} {arXiv}\ }\textbf {\bibinfo {volume} {2307.08937}} (\bibinfo {year} {2023})}\BibitemShut {NoStop}%
\bibitem [{\citenamefont {Sun}\ \emph {et~al.}(2023)\citenamefont {Sun}, \citenamefont {Wang}, \citenamefont {Liu}, \citenamefont {Harrington}, \citenamefont {Tabatabaei}, \citenamefont {Liu}, \citenamefont {Wang}, \citenamefont {Hanifi}, \citenamefont {Morgan}, \citenamefont {Jahanbozorgi}, \citenamefont {Yang}, \citenamefont {Bowers}, \citenamefont {Morton}, \citenamefont {Nelson}, \citenamefont {Beling}, \citenamefont {Blumenthal},\ and\ \citenamefont {Yi}}]{Sun:23}%
  \BibitemOpen
  \bibfield  {author} {\bibinfo {author} {\bibfnamefont {S.}~\bibnamefont {Sun}}, \bibinfo {author} {\bibfnamefont {B.}~\bibnamefont {Wang}}, \bibinfo {author} {\bibfnamefont {K.}~\bibnamefont {Liu}}, \bibinfo {author} {\bibfnamefont {M.}~\bibnamefont {Harrington}}, \bibinfo {author} {\bibfnamefont {F.}~\bibnamefont {Tabatabaei}}, \bibinfo {author} {\bibfnamefont {R.}~\bibnamefont {Liu}}, \bibinfo {author} {\bibfnamefont {J.}~\bibnamefont {Wang}}, \bibinfo {author} {\bibfnamefont {S.}~\bibnamefont {Hanifi}}, \bibinfo {author} {\bibfnamefont {J.~S.}\ \bibnamefont {Morgan}}, \bibinfo {author} {\bibfnamefont {M.}~\bibnamefont {Jahanbozorgi}}, \bibinfo {author} {\bibfnamefont {Z.}~\bibnamefont {Yang}}, \bibinfo {author} {\bibfnamefont {S.}~\bibnamefont {Bowers}}, \bibinfo {author} {\bibfnamefont {P.}~\bibnamefont {Morton}}, \bibinfo {author} {\bibfnamefont {K.}~\bibnamefont {Nelson}}, \bibinfo {author} {\bibfnamefont {A.}~\bibnamefont {Beling}}, \bibinfo {author} {\bibfnamefont {D.}~\bibnamefont {Blumenthal}}, \
  and\ \bibinfo {author} {\bibfnamefont {X.}~\bibnamefont {Yi}},\ }\href@noop {} {\bibfield  {journal} {\bibinfo  {journal} {arXiv}\ }\textbf {\bibinfo {volume} {2305.13575}} (\bibinfo {year} {2023})}\BibitemShut {NoStop}%
\bibitem [{\citenamefont {Jin}\ \emph {et~al.}(2024)\citenamefont {Jin}, \citenamefont {Xie}, \citenamefont {Zhang}, \citenamefont {Hou}, \citenamefont {Zhang}, \citenamefont {Zhang}, \citenamefont {Gong}, \citenamefont {Chang},\ and\ \citenamefont {Yang}}]{Jin:24}%
  \BibitemOpen
  \bibfield  {author} {\bibinfo {author} {\bibfnamefont {X.}~\bibnamefont {Jin}}, \bibinfo {author} {\bibfnamefont {Z.}~\bibnamefont {Xie}}, \bibinfo {author} {\bibfnamefont {X.}~\bibnamefont {Zhang}}, \bibinfo {author} {\bibfnamefont {H.}~\bibnamefont {Hou}}, \bibinfo {author} {\bibfnamefont {F.}~\bibnamefont {Zhang}}, \bibinfo {author} {\bibfnamefont {X.}~\bibnamefont {Zhang}}, \bibinfo {author} {\bibfnamefont {Q.}~\bibnamefont {Gong}}, \bibinfo {author} {\bibfnamefont {L.}~\bibnamefont {Chang}}, \ and\ \bibinfo {author} {\bibfnamefont {Q.-F.}\ \bibnamefont {Yang}},\ }\href@noop {} {\bibfield  {journal} {\bibinfo  {journal} {arXiv}\ }\textbf {\bibinfo {volume} {2401.12760}} (\bibinfo {year} {2024})}\BibitemShut {NoStop}%
\bibitem [{\citenamefont {He}\ \emph {et~al.}(2024)\citenamefont {He}, \citenamefont {Cheng}, \citenamefont {Wang}, \citenamefont {Zhang}, \citenamefont {Meade}, \citenamefont {Vahala}, \citenamefont {Zhang},\ and\ \citenamefont {Li}}]{He:24}%
  \BibitemOpen
  \bibfield  {author} {\bibinfo {author} {\bibfnamefont {Y.}~\bibnamefont {He}}, \bibinfo {author} {\bibfnamefont {L.}~\bibnamefont {Cheng}}, \bibinfo {author} {\bibfnamefont {H.}~\bibnamefont {Wang}}, \bibinfo {author} {\bibfnamefont {Y.}~\bibnamefont {Zhang}}, \bibinfo {author} {\bibfnamefont {R.}~\bibnamefont {Meade}}, \bibinfo {author} {\bibfnamefont {K.}~\bibnamefont {Vahala}}, \bibinfo {author} {\bibfnamefont {M.}~\bibnamefont {Zhang}}, \ and\ \bibinfo {author} {\bibfnamefont {J.}~\bibnamefont {Li}},\ }\href@noop {} {\bibfield  {journal} {\bibinfo  {journal} {arXiv}\ }\textbf {\bibinfo {volume} {2402.16229}} (\bibinfo {year} {2024})}\BibitemShut {NoStop}%
\bibitem [{\citenamefont {Stern}\ \emph {et~al.}(2018)\citenamefont {Stern}, \citenamefont {Ji}, \citenamefont {Okawachi}, \citenamefont {Gaeta},\ and\ \citenamefont {Lipson}}]{Stern:18}%
  \BibitemOpen
  \bibfield  {author} {\bibinfo {author} {\bibfnamefont {B.}~\bibnamefont {Stern}}, \bibinfo {author} {\bibfnamefont {X.}~\bibnamefont {Ji}}, \bibinfo {author} {\bibfnamefont {Y.}~\bibnamefont {Okawachi}}, \bibinfo {author} {\bibfnamefont {A.~L.}\ \bibnamefont {Gaeta}}, \ and\ \bibinfo {author} {\bibfnamefont {M.}~\bibnamefont {Lipson}},\ }\href {\doibase 10.1038/s41586-018-0598-9} {\bibfield  {journal} {\bibinfo  {journal} {Nature}\ }\textbf {\bibinfo {volume} {562}},\ \bibinfo {pages} {401} (\bibinfo {year} {2018})}\BibitemShut {NoStop}%
\bibitem [{\citenamefont {Raja}\ \emph {et~al.}(2019)\citenamefont {Raja}, \citenamefont {Voloshin}, \citenamefont {Guo}, \citenamefont {Agafonova}, \citenamefont {Liu}, \citenamefont {Gorodnitskiy}, \citenamefont {Karpov}, \citenamefont {Pavlov}, \citenamefont {Lucas}, \citenamefont {Galiev}, \citenamefont {Shitikov}, \citenamefont {Jost}, \citenamefont {Gorodetsky},\ and\ \citenamefont {Kippenberg}}]{Raja:19}%
  \BibitemOpen
  \bibfield  {author} {\bibinfo {author} {\bibfnamefont {A.~S.}\ \bibnamefont {Raja}}, \bibinfo {author} {\bibfnamefont {A.~S.}\ \bibnamefont {Voloshin}}, \bibinfo {author} {\bibfnamefont {H.}~\bibnamefont {Guo}}, \bibinfo {author} {\bibfnamefont {S.~E.}\ \bibnamefont {Agafonova}}, \bibinfo {author} {\bibfnamefont {J.}~\bibnamefont {Liu}}, \bibinfo {author} {\bibfnamefont {A.~S.}\ \bibnamefont {Gorodnitskiy}}, \bibinfo {author} {\bibfnamefont {M.}~\bibnamefont {Karpov}}, \bibinfo {author} {\bibfnamefont {N.~G.}\ \bibnamefont {Pavlov}}, \bibinfo {author} {\bibfnamefont {E.}~\bibnamefont {Lucas}}, \bibinfo {author} {\bibfnamefont {R.~R.}\ \bibnamefont {Galiev}}, \bibinfo {author} {\bibfnamefont {A.~E.}\ \bibnamefont {Shitikov}}, \bibinfo {author} {\bibfnamefont {J.~D.}\ \bibnamefont {Jost}}, \bibinfo {author} {\bibfnamefont {M.~L.}\ \bibnamefont {Gorodetsky}}, \ and\ \bibinfo {author} {\bibfnamefont {T.~J.}\ \bibnamefont {Kippenberg}},\ }\href {\doibase 10.1038/s41467-019-08498-2} {\bibfield  {journal}
  {\bibinfo  {journal} {Nature Communications}\ }\textbf {\bibinfo {volume} {10}},\ \bibinfo {pages} {680} (\bibinfo {year} {2019})}\BibitemShut {NoStop}%
\bibitem [{\citenamefont {Shen}\ \emph {et~al.}(2020)\citenamefont {Shen}, \citenamefont {Chang}, \citenamefont {Liu}, \citenamefont {Wang}, \citenamefont {Yang}, \citenamefont {Xiang}, \citenamefont {Wang}, \citenamefont {He}, \citenamefont {Liu}, \citenamefont {Xie}, \citenamefont {Guo}, \citenamefont {Kinghorn}, \citenamefont {Wu}, \citenamefont {Ji}, \citenamefont {Kippenberg}, \citenamefont {Vahala},\ and\ \citenamefont {Bowers}}]{Shen:20}%
  \BibitemOpen
  \bibfield  {author} {\bibinfo {author} {\bibfnamefont {B.}~\bibnamefont {Shen}}, \bibinfo {author} {\bibfnamefont {L.}~\bibnamefont {Chang}}, \bibinfo {author} {\bibfnamefont {J.}~\bibnamefont {Liu}}, \bibinfo {author} {\bibfnamefont {H.}~\bibnamefont {Wang}}, \bibinfo {author} {\bibfnamefont {Q.-F.}\ \bibnamefont {Yang}}, \bibinfo {author} {\bibfnamefont {C.}~\bibnamefont {Xiang}}, \bibinfo {author} {\bibfnamefont {R.~N.}\ \bibnamefont {Wang}}, \bibinfo {author} {\bibfnamefont {J.}~\bibnamefont {He}}, \bibinfo {author} {\bibfnamefont {T.}~\bibnamefont {Liu}}, \bibinfo {author} {\bibfnamefont {W.}~\bibnamefont {Xie}}, \bibinfo {author} {\bibfnamefont {J.}~\bibnamefont {Guo}}, \bibinfo {author} {\bibfnamefont {D.}~\bibnamefont {Kinghorn}}, \bibinfo {author} {\bibfnamefont {L.}~\bibnamefont {Wu}}, \bibinfo {author} {\bibfnamefont {Q.-X.}\ \bibnamefont {Ji}}, \bibinfo {author} {\bibfnamefont {T.~J.}\ \bibnamefont {Kippenberg}}, \bibinfo {author} {\bibfnamefont {K.}~\bibnamefont {Vahala}}, \ and\ \bibinfo
  {author} {\bibfnamefont {J.~E.}\ \bibnamefont {Bowers}},\ }\href {\doibase 10.1038/s41586-020-2358-x} {\bibfield  {journal} {\bibinfo  {journal} {Nature}\ }\textbf {\bibinfo {volume} {582}},\ \bibinfo {pages} {365} (\bibinfo {year} {2020})}\BibitemShut {NoStop}%
\bibitem [{\citenamefont {Lihachev}\ \emph {et~al.}(2022)\citenamefont {Lihachev}, \citenamefont {Weng}, \citenamefont {Liu}, \citenamefont {Chang}, \citenamefont {Guo}, \citenamefont {He}, \citenamefont {Wang}, \citenamefont {Anderson}, \citenamefont {Liu}, \citenamefont {Bowers},\ and\ \citenamefont {Kippenberg}}]{Lihachev:22a}%
  \BibitemOpen
  \bibfield  {author} {\bibinfo {author} {\bibfnamefont {G.}~\bibnamefont {Lihachev}}, \bibinfo {author} {\bibfnamefont {W.}~\bibnamefont {Weng}}, \bibinfo {author} {\bibfnamefont {J.}~\bibnamefont {Liu}}, \bibinfo {author} {\bibfnamefont {L.}~\bibnamefont {Chang}}, \bibinfo {author} {\bibfnamefont {J.}~\bibnamefont {Guo}}, \bibinfo {author} {\bibfnamefont {J.}~\bibnamefont {He}}, \bibinfo {author} {\bibfnamefont {R.~N.}\ \bibnamefont {Wang}}, \bibinfo {author} {\bibfnamefont {M.~H.}\ \bibnamefont {Anderson}}, \bibinfo {author} {\bibfnamefont {Y.}~\bibnamefont {Liu}}, \bibinfo {author} {\bibfnamefont {J.~E.}\ \bibnamefont {Bowers}}, \ and\ \bibinfo {author} {\bibfnamefont {T.~J.}\ \bibnamefont {Kippenberg}},\ }\href {\doibase 10.1038/s41467-022-29431-0} {\bibfield  {journal} {\bibinfo  {journal} {Nature Communications}\ }\textbf {\bibinfo {volume} {13}},\ \bibinfo {pages} {1771} (\bibinfo {year} {2022})}\BibitemShut {NoStop}%
\bibitem [{\citenamefont {Ye}\ \emph {et~al.}(2023)\citenamefont {Ye}, \citenamefont {Jia}, \citenamefont {Huang}, \citenamefont {Shen}, \citenamefont {Long}, \citenamefont {Shi}, \citenamefont {Luo}, \citenamefont {Gao}, \citenamefont {Sun}, \citenamefont {Guo}, \citenamefont {He},\ and\ \citenamefont {Liu}}]{Ye:23}%
  \BibitemOpen
  \bibfield  {author} {\bibinfo {author} {\bibfnamefont {Z.}~\bibnamefont {Ye}}, \bibinfo {author} {\bibfnamefont {H.}~\bibnamefont {Jia}}, \bibinfo {author} {\bibfnamefont {Z.}~\bibnamefont {Huang}}, \bibinfo {author} {\bibfnamefont {C.}~\bibnamefont {Shen}}, \bibinfo {author} {\bibfnamefont {J.}~\bibnamefont {Long}}, \bibinfo {author} {\bibfnamefont {B.}~\bibnamefont {Shi}}, \bibinfo {author} {\bibfnamefont {Y.-H.}\ \bibnamefont {Luo}}, \bibinfo {author} {\bibfnamefont {L.}~\bibnamefont {Gao}}, \bibinfo {author} {\bibfnamefont {W.}~\bibnamefont {Sun}}, \bibinfo {author} {\bibfnamefont {H.}~\bibnamefont {Guo}}, \bibinfo {author} {\bibfnamefont {J.}~\bibnamefont {He}}, \ and\ \bibinfo {author} {\bibfnamefont {J.}~\bibnamefont {Liu}},\ }\href {\doibase 10.1364/PRJ.486379} {\bibfield  {journal} {\bibinfo  {journal} {Photon. Res.}\ }\textbf {\bibinfo {volume} {11}},\ \bibinfo {pages} {558} (\bibinfo {year} {2023})}\BibitemShut {NoStop}%
\bibitem [{\citenamefont {Luke}\ \emph {et~al.}(2013)\citenamefont {Luke}, \citenamefont {Dutt}, \citenamefont {Poitras},\ and\ \citenamefont {Lipson}}]{Luke:13}%
  \BibitemOpen
  \bibfield  {author} {\bibinfo {author} {\bibfnamefont {K.}~\bibnamefont {Luke}}, \bibinfo {author} {\bibfnamefont {A.}~\bibnamefont {Dutt}}, \bibinfo {author} {\bibfnamefont {C.~B.}\ \bibnamefont {Poitras}}, \ and\ \bibinfo {author} {\bibfnamefont {M.}~\bibnamefont {Lipson}},\ }\href {\doibase 10.1364/OE.21.022829} {\bibfield  {journal} {\bibinfo  {journal} {Opt. Express}\ }\textbf {\bibinfo {volume} {21}},\ \bibinfo {pages} {22829} (\bibinfo {year} {2013})}\BibitemShut {NoStop}%
\bibitem [{\citenamefont {Mu{\~n}oz}\ \emph {et~al.}(2019)\citenamefont {Mu{\~n}oz}, \citenamefont {van Dijk}, \citenamefont {Geuzebroek}, \citenamefont {Geiselmann}, \citenamefont {Dom{\'\i}nguez}, \citenamefont {Stassen}, \citenamefont {Dom{\'e}nech}, \citenamefont {Zervas}, \citenamefont {Leinse}, \citenamefont {Roeloffzen}, \citenamefont {Gargallo}, \citenamefont {Ba{\~n}os}, \citenamefont {Fern{\'a}ndez}, \citenamefont {Cabanes}, \citenamefont {Bru},\ and\ \citenamefont {Pastor}}]{Munoz:19}%
  \BibitemOpen
  \bibfield  {author} {\bibinfo {author} {\bibfnamefont {P.}~\bibnamefont {Mu{\~n}oz}}, \bibinfo {author} {\bibfnamefont {P.~W.~L.}\ \bibnamefont {van Dijk}}, \bibinfo {author} {\bibfnamefont {D.}~\bibnamefont {Geuzebroek}}, \bibinfo {author} {\bibfnamefont {M.}~\bibnamefont {Geiselmann}}, \bibinfo {author} {\bibfnamefont {C.}~\bibnamefont {Dom{\'\i}nguez}}, \bibinfo {author} {\bibfnamefont {A.}~\bibnamefont {Stassen}}, \bibinfo {author} {\bibfnamefont {J.~D.}\ \bibnamefont {Dom{\'e}nech}}, \bibinfo {author} {\bibfnamefont {M.}~\bibnamefont {Zervas}}, \bibinfo {author} {\bibfnamefont {A.}~\bibnamefont {Leinse}}, \bibinfo {author} {\bibfnamefont {C.~G.~H.}\ \bibnamefont {Roeloffzen}}, \bibinfo {author} {\bibfnamefont {B.}~\bibnamefont {Gargallo}}, \bibinfo {author} {\bibfnamefont {R.}~\bibnamefont {Ba{\~n}os}}, \bibinfo {author} {\bibfnamefont {J.}~\bibnamefont {Fern{\'a}ndez}}, \bibinfo {author} {\bibfnamefont {G.~M.}\ \bibnamefont {Cabanes}}, \bibinfo {author} {\bibfnamefont {L.~A.}\ \bibnamefont {Bru}}, \
  and\ \bibinfo {author} {\bibfnamefont {D.}~\bibnamefont {Pastor}},\ }\href {\doibase 10.1109/JSTQE.2019.2902903} {\bibfield  {journal} {\bibinfo  {journal} {IEEE Journal of Selected Topics in Quantum Electronics}\ }\textbf {\bibinfo {volume} {25}},\ \bibinfo {pages} {1} (\bibinfo {year} {2019})}\BibitemShut {NoStop}%
\bibitem [{\citenamefont {Xiang}\ \emph {et~al.}(2022)\citenamefont {Xiang}, \citenamefont {Jin},\ and\ \citenamefont {Bowers}}]{Xiang:22a}%
  \BibitemOpen
  \bibfield  {author} {\bibinfo {author} {\bibfnamefont {C.}~\bibnamefont {Xiang}}, \bibinfo {author} {\bibfnamefont {W.}~\bibnamefont {Jin}}, \ and\ \bibinfo {author} {\bibfnamefont {J.~E.}\ \bibnamefont {Bowers}},\ }\href {\doibase 10.1364/PRJ.452936} {\bibfield  {journal} {\bibinfo  {journal} {Photon. Res.}\ }\textbf {\bibinfo {volume} {10}},\ \bibinfo {pages} {A82} (\bibinfo {year} {2022})}\BibitemShut {NoStop}%
\bibitem [{\citenamefont {Luo}\ \emph {et~al.}(2023)\citenamefont {Luo}, \citenamefont {Shi}, \citenamefont {Sun}, \citenamefont {Chen}, \citenamefont {Huang}, \citenamefont {Wang}, \citenamefont {Long}, \citenamefont {Shen}, \citenamefont {Ye}, \citenamefont {Guo},\ and\ \citenamefont {Liu}}]{Luo:23}%
  \BibitemOpen
  \bibfield  {author} {\bibinfo {author} {\bibfnamefont {Y.-H.}\ \bibnamefont {Luo}}, \bibinfo {author} {\bibfnamefont {B.}~\bibnamefont {Shi}}, \bibinfo {author} {\bibfnamefont {W.}~\bibnamefont {Sun}}, \bibinfo {author} {\bibfnamefont {R.}~\bibnamefont {Chen}}, \bibinfo {author} {\bibfnamefont {S.}~\bibnamefont {Huang}}, \bibinfo {author} {\bibfnamefont {Z.}~\bibnamefont {Wang}}, \bibinfo {author} {\bibfnamefont {J.}~\bibnamefont {Long}}, \bibinfo {author} {\bibfnamefont {C.}~\bibnamefont {Shen}}, \bibinfo {author} {\bibfnamefont {Z.}~\bibnamefont {Ye}}, \bibinfo {author} {\bibfnamefont {H.}~\bibnamefont {Guo}}, \ and\ \bibinfo {author} {\bibfnamefont {J.}~\bibnamefont {Liu}},\ }\href@noop {} {\bibfield  {journal} {\bibinfo  {journal} {arXiv}\ }\textbf {\bibinfo {volume} {2304.04295}} (\bibinfo {year} {2023})}\BibitemShut {NoStop}%
\bibitem [{\citenamefont {Lobanov}\ \emph {et~al.}(2015)\citenamefont {Lobanov}, \citenamefont {Lihachev}, \citenamefont {Kippenberg},\ and\ \citenamefont {Gorodetsky}}]{Lobanov:15}%
  \BibitemOpen
  \bibfield  {author} {\bibinfo {author} {\bibfnamefont {V.}~\bibnamefont {Lobanov}}, \bibinfo {author} {\bibfnamefont {G.}~\bibnamefont {Lihachev}}, \bibinfo {author} {\bibfnamefont {T.~J.}\ \bibnamefont {Kippenberg}}, \ and\ \bibinfo {author} {\bibfnamefont {M.}~\bibnamefont {Gorodetsky}},\ }\href {\doibase 10.1364/OE.23.007713} {\bibfield  {journal} {\bibinfo  {journal} {Opt. Express}\ }\textbf {\bibinfo {volume} {23}},\ \bibinfo {pages} {7713} (\bibinfo {year} {2015})}\BibitemShut {NoStop}%
\bibitem [{\citenamefont {Huang}\ \emph {et~al.}(2015)\citenamefont {Huang}, \citenamefont {Zhou}, \citenamefont {Yang}, \citenamefont {McMillan}, \citenamefont {Matsko}, \citenamefont {Yu}, \citenamefont {Kwong}, \citenamefont {Maleki},\ and\ \citenamefont {Wong}}]{Huang:15b}%
  \BibitemOpen
  \bibfield  {author} {\bibinfo {author} {\bibfnamefont {S.-W.}\ \bibnamefont {Huang}}, \bibinfo {author} {\bibfnamefont {H.}~\bibnamefont {Zhou}}, \bibinfo {author} {\bibfnamefont {J.}~\bibnamefont {Yang}}, \bibinfo {author} {\bibfnamefont {J.~F.}\ \bibnamefont {McMillan}}, \bibinfo {author} {\bibfnamefont {A.}~\bibnamefont {Matsko}}, \bibinfo {author} {\bibfnamefont {M.}~\bibnamefont {Yu}}, \bibinfo {author} {\bibfnamefont {D.-L.}\ \bibnamefont {Kwong}}, \bibinfo {author} {\bibfnamefont {L.}~\bibnamefont {Maleki}}, \ and\ \bibinfo {author} {\bibfnamefont {C.~W.}\ \bibnamefont {Wong}},\ }\href {\doibase 10.1103/PhysRevLett.114.053901} {\bibfield  {journal} {\bibinfo  {journal} {Phys. Rev. Lett.}\ }\textbf {\bibinfo {volume} {114}},\ \bibinfo {pages} {053901} (\bibinfo {year} {2015})}\BibitemShut {NoStop}%
\bibitem [{\citenamefont {Parra-Rivas}\ \emph {et~al.}(2016)\citenamefont {Parra-Rivas}, \citenamefont {Gomila}, \citenamefont {Knobloch}, \citenamefont {Coen},\ and\ \citenamefont {Gelens}}]{Parra-Rivas:16}%
  \BibitemOpen
  \bibfield  {author} {\bibinfo {author} {\bibfnamefont {P.}~\bibnamefont {Parra-Rivas}}, \bibinfo {author} {\bibfnamefont {D.}~\bibnamefont {Gomila}}, \bibinfo {author} {\bibfnamefont {E.}~\bibnamefont {Knobloch}}, \bibinfo {author} {\bibfnamefont {S.}~\bibnamefont {Coen}}, \ and\ \bibinfo {author} {\bibfnamefont {L.}~\bibnamefont {Gelens}},\ }\href {\doibase 10.1364/OL.41.002402} {\bibfield  {journal} {\bibinfo  {journal} {Optics Letters}\ }\textbf {\bibinfo {volume} {41}},\ \bibinfo {pages} {2402} (\bibinfo {year} {2016})}\BibitemShut {NoStop}%
\bibitem [{\citenamefont {Nazemosadat}\ \emph {et~al.}(2021)\citenamefont {Nazemosadat}, \citenamefont {F\"ul\"op}, \citenamefont {Helgason}, \citenamefont {Wang}, \citenamefont {Xuan}, \citenamefont {Leaird}, \citenamefont {Qi}, \citenamefont {Silvestre}, \citenamefont {Weiner},\ and\ \citenamefont {Torres-Company}}]{Nazemosadat:21}%
  \BibitemOpen
  \bibfield  {author} {\bibinfo {author} {\bibfnamefont {E.}~\bibnamefont {Nazemosadat}}, \bibinfo {author} {\bibfnamefont {A.}~\bibnamefont {F\"ul\"op}}, \bibinfo {author} {\bibfnamefont {O.~B.}\ \bibnamefont {Helgason}}, \bibinfo {author} {\bibfnamefont {P.-H.}\ \bibnamefont {Wang}}, \bibinfo {author} {\bibfnamefont {Y.}~\bibnamefont {Xuan}}, \bibinfo {author} {\bibfnamefont {D.~E.}\ \bibnamefont {Leaird}}, \bibinfo {author} {\bibfnamefont {M.}~\bibnamefont {Qi}}, \bibinfo {author} {\bibfnamefont {E.}~\bibnamefont {Silvestre}}, \bibinfo {author} {\bibfnamefont {A.~M.}\ \bibnamefont {Weiner}}, \ and\ \bibinfo {author} {\bibfnamefont {V.}~\bibnamefont {Torres-Company}},\ }\href {\doibase 10.1103/PhysRevA.103.013513} {\bibfield  {journal} {\bibinfo  {journal} {Phys. Rev. A}\ }\textbf {\bibinfo {volume} {103}},\ \bibinfo {pages} {013513} (\bibinfo {year} {2021})}\BibitemShut {NoStop}%
\bibitem [{\citenamefont {Okawachi}\ \emph {et~al.}(2014)\citenamefont {Okawachi}, \citenamefont {Lamont}, \citenamefont {Luke}, \citenamefont {Carvalho}, \citenamefont {Yu}, \citenamefont {Lipson},\ and\ \citenamefont {Gaeta}}]{Okawachi:14}%
  \BibitemOpen
  \bibfield  {author} {\bibinfo {author} {\bibfnamefont {Y.}~\bibnamefont {Okawachi}}, \bibinfo {author} {\bibfnamefont {M.~R.~E.}\ \bibnamefont {Lamont}}, \bibinfo {author} {\bibfnamefont {K.}~\bibnamefont {Luke}}, \bibinfo {author} {\bibfnamefont {D.~O.}\ \bibnamefont {Carvalho}}, \bibinfo {author} {\bibfnamefont {M.}~\bibnamefont {Yu}}, \bibinfo {author} {\bibfnamefont {M.}~\bibnamefont {Lipson}}, \ and\ \bibinfo {author} {\bibfnamefont {A.~L.}\ \bibnamefont {Gaeta}},\ }\href {\doibase 10.1364/OL.39.003535} {\bibfield  {journal} {\bibinfo  {journal} {Opt. Lett.}\ }\textbf {\bibinfo {volume} {39}},\ \bibinfo {pages} {3535} (\bibinfo {year} {2014})}\BibitemShut {NoStop}%
\bibitem [{\citenamefont {Xue}\ \emph {et~al.}(2017)\citenamefont {Xue}, \citenamefont {Wang}, \citenamefont {Xuan}, \citenamefont {Qi},\ and\ \citenamefont {Weiner}}]{Xue:17b}%
  \BibitemOpen
  \bibfield  {author} {\bibinfo {author} {\bibfnamefont {X.}~\bibnamefont {Xue}}, \bibinfo {author} {\bibfnamefont {P.-H.}\ \bibnamefont {Wang}}, \bibinfo {author} {\bibfnamefont {Y.}~\bibnamefont {Xuan}}, \bibinfo {author} {\bibfnamefont {M.}~\bibnamefont {Qi}}, \ and\ \bibinfo {author} {\bibfnamefont {A.~M.}\ \bibnamefont {Weiner}},\ }\href {\doibase https://doi.org/10.1002/lpor.201600276} {\bibfield  {journal} {\bibinfo  {journal} {Laser \& Photonics Reviews}\ }\textbf {\bibinfo {volume} {11}},\ \bibinfo {pages} {1600276} (\bibinfo {year} {2017})}\BibitemShut {NoStop}%
\bibitem [{\citenamefont {Jang}\ \emph {et~al.}(2021)\citenamefont {Jang}, \citenamefont {Okawachi}, \citenamefont {Zhao}, \citenamefont {Ji}, \citenamefont {Joshi}, \citenamefont {Lipson},\ and\ \citenamefont {Gaeta}}]{JangJ:21}%
  \BibitemOpen
  \bibfield  {author} {\bibinfo {author} {\bibfnamefont {J.~K.}\ \bibnamefont {Jang}}, \bibinfo {author} {\bibfnamefont {Y.}~\bibnamefont {Okawachi}}, \bibinfo {author} {\bibfnamefont {Y.}~\bibnamefont {Zhao}}, \bibinfo {author} {\bibfnamefont {X.}~\bibnamefont {Ji}}, \bibinfo {author} {\bibfnamefont {C.}~\bibnamefont {Joshi}}, \bibinfo {author} {\bibfnamefont {M.}~\bibnamefont {Lipson}}, \ and\ \bibinfo {author} {\bibfnamefont {A.~L.}\ \bibnamefont {Gaeta}},\ }\href {\doibase 10.1364/OL.423654} {\bibfield  {journal} {\bibinfo  {journal} {Opt. Lett.}\ }\textbf {\bibinfo {volume} {46}},\ \bibinfo {pages} {3657} (\bibinfo {year} {2021})}\BibitemShut {NoStop}%
\bibitem [{\citenamefont {Li}\ \emph {et~al.}(2023)\citenamefont {Li}, \citenamefont {Wang},\ and\ \citenamefont {Chen}}]{Li:23}%
  \BibitemOpen
  \bibfield  {author} {\bibinfo {author} {\bibfnamefont {L.}~\bibnamefont {Li}}, \bibinfo {author} {\bibfnamefont {L.}~\bibnamefont {Wang}}, \ and\ \bibinfo {author} {\bibfnamefont {B.}~\bibnamefont {Chen}},\ }in\ \href {\doibase 10.1109/OECC56963.2023.10209607} {\emph {\bibinfo {booktitle} {2023 Opto-Electronics and Communications Conference (OECC)}}}\ (\bibinfo {year} {2023})\ pp.\ \bibinfo {pages} {1--3}\BibitemShut {NoStop}%
\bibitem [{\citenamefont {Jin}\ \emph {et~al.}(2021)\citenamefont {Jin}, \citenamefont {Yang}, \citenamefont {Chang}, \citenamefont {Shen}, \citenamefont {Wang}, \citenamefont {Leal}, \citenamefont {Wu}, \citenamefont {Gao}, \citenamefont {Feshali}, \citenamefont {Paniccia}, \citenamefont {Vahala},\ and\ \citenamefont {Bowers}}]{Jin:21}%
  \BibitemOpen
  \bibfield  {author} {\bibinfo {author} {\bibfnamefont {W.}~\bibnamefont {Jin}}, \bibinfo {author} {\bibfnamefont {Q.-F.}\ \bibnamefont {Yang}}, \bibinfo {author} {\bibfnamefont {L.}~\bibnamefont {Chang}}, \bibinfo {author} {\bibfnamefont {B.}~\bibnamefont {Shen}}, \bibinfo {author} {\bibfnamefont {H.}~\bibnamefont {Wang}}, \bibinfo {author} {\bibfnamefont {M.~A.}\ \bibnamefont {Leal}}, \bibinfo {author} {\bibfnamefont {L.}~\bibnamefont {Wu}}, \bibinfo {author} {\bibfnamefont {M.}~\bibnamefont {Gao}}, \bibinfo {author} {\bibfnamefont {A.}~\bibnamefont {Feshali}}, \bibinfo {author} {\bibfnamefont {M.}~\bibnamefont {Paniccia}}, \bibinfo {author} {\bibfnamefont {K.~J.}\ \bibnamefont {Vahala}}, \ and\ \bibinfo {author} {\bibfnamefont {J.~E.}\ \bibnamefont {Bowers}},\ }\href {\doibase 10.1038/s41566-021-00761-7} {\bibfield  {journal} {\bibinfo  {journal} {Nature Photonics}\ }\textbf {\bibinfo {volume} {15}},\ \bibinfo {pages} {346} (\bibinfo {year} {2021})}\BibitemShut {NoStop}%
\bibitem [{\citenamefont {Liang}\ \emph {et~al.}(2015{\natexlab{b}})\citenamefont {Liang}, \citenamefont {Ilchenko}, \citenamefont {Eliyahu}, \citenamefont {Savchenkov}, \citenamefont {Matsko}, \citenamefont {Seidel},\ and\ \citenamefont {Maleki}}]{Liang:15b}%
  \BibitemOpen
  \bibfield  {author} {\bibinfo {author} {\bibfnamefont {W.}~\bibnamefont {Liang}}, \bibinfo {author} {\bibfnamefont {V.~S.}\ \bibnamefont {Ilchenko}}, \bibinfo {author} {\bibfnamefont {D.}~\bibnamefont {Eliyahu}}, \bibinfo {author} {\bibfnamefont {A.~A.}\ \bibnamefont {Savchenkov}}, \bibinfo {author} {\bibfnamefont {A.~B.}\ \bibnamefont {Matsko}}, \bibinfo {author} {\bibfnamefont {D.}~\bibnamefont {Seidel}}, \ and\ \bibinfo {author} {\bibfnamefont {L.}~\bibnamefont {Maleki}},\ }\href {\doibase 10.1038/ncomms8371} {\bibfield  {journal} {\bibinfo  {journal} {Nature Communications}\ }\textbf {\bibinfo {volume} {6}},\ \bibinfo {pages} {7371} (\bibinfo {year} {2015}{\natexlab{b}})}\BibitemShut {NoStop}%
\bibitem [{\citenamefont {Kondratiev}\ \emph {et~al.}(2017)\citenamefont {Kondratiev}, \citenamefont {Lobanov}, \citenamefont {Cherenkov}, \citenamefont {Voloshin}, \citenamefont {Pavlov}, \citenamefont {Koptyaev},\ and\ \citenamefont {Gorodetsky}}]{Kondratiev:17}%
  \BibitemOpen
  \bibfield  {author} {\bibinfo {author} {\bibfnamefont {N.~M.}\ \bibnamefont {Kondratiev}}, \bibinfo {author} {\bibfnamefont {V.~E.}\ \bibnamefont {Lobanov}}, \bibinfo {author} {\bibfnamefont {A.~V.}\ \bibnamefont {Cherenkov}}, \bibinfo {author} {\bibfnamefont {A.~S.}\ \bibnamefont {Voloshin}}, \bibinfo {author} {\bibfnamefont {N.~G.}\ \bibnamefont {Pavlov}}, \bibinfo {author} {\bibfnamefont {S.}~\bibnamefont {Koptyaev}}, \ and\ \bibinfo {author} {\bibfnamefont {M.~L.}\ \bibnamefont {Gorodetsky}},\ }\href {\doibase 10.1364/OE.25.028167} {\bibfield  {journal} {\bibinfo  {journal} {Optics Express}\ }\textbf {\bibinfo {volume} {25}},\ \bibinfo {pages} {28167} (\bibinfo {year} {2017})}\BibitemShut {NoStop}%
\bibitem [{\citenamefont {Kondratiev}\ \emph {et~al.}(2023)\citenamefont {Kondratiev}, \citenamefont {Lobanov}, \citenamefont {Shitikov}, \citenamefont {Galiev}, \citenamefont {Chermoshentsev}, \citenamefont {Dmitriev}, \citenamefont {Danilin}, \citenamefont {Lonshakov}, \citenamefont {Min'kov}, \citenamefont {Sokol}, \citenamefont {Cordette}, \citenamefont {Luo}, \citenamefont {Liang}, \citenamefont {Liu},\ and\ \citenamefont {Bilenko}}]{Kondratiev:23}%
  \BibitemOpen
  \bibfield  {author} {\bibinfo {author} {\bibfnamefont {N.~M.}\ \bibnamefont {Kondratiev}}, \bibinfo {author} {\bibfnamefont {V.~E.}\ \bibnamefont {Lobanov}}, \bibinfo {author} {\bibfnamefont {A.~E.}\ \bibnamefont {Shitikov}}, \bibinfo {author} {\bibfnamefont {R.~R.}\ \bibnamefont {Galiev}}, \bibinfo {author} {\bibfnamefont {D.~A.}\ \bibnamefont {Chermoshentsev}}, \bibinfo {author} {\bibfnamefont {N.~Y.}\ \bibnamefont {Dmitriev}}, \bibinfo {author} {\bibfnamefont {A.~N.}\ \bibnamefont {Danilin}}, \bibinfo {author} {\bibfnamefont {E.~A.}\ \bibnamefont {Lonshakov}}, \bibinfo {author} {\bibfnamefont {K.~N.}\ \bibnamefont {Min'kov}}, \bibinfo {author} {\bibfnamefont {D.~M.}\ \bibnamefont {Sokol}}, \bibinfo {author} {\bibfnamefont {S.~J.}\ \bibnamefont {Cordette}}, \bibinfo {author} {\bibfnamefont {Y.-H.}\ \bibnamefont {Luo}}, \bibinfo {author} {\bibfnamefont {W.}~\bibnamefont {Liang}}, \bibinfo {author} {\bibfnamefont {J.}~\bibnamefont {Liu}}, \ and\ \bibinfo {author} {\bibfnamefont {I.~A.}\ \bibnamefont
  {Bilenko}},\ }\href {\doibase 10.1007/s11467-022-1245-3} {\bibfield  {journal} {\bibinfo  {journal} {Frontiers of Physics}\ }\textbf {\bibinfo {volume} {18}},\ \bibinfo {pages} {21305} (\bibinfo {year} {2023})}\BibitemShut {NoStop}%
\bibitem [{\citenamefont {Voloshin}\ \emph {et~al.}(2021)\citenamefont {Voloshin}, \citenamefont {Kondratiev}, \citenamefont {Lihachev}, \citenamefont {Liu}, \citenamefont {Lobanov}, \citenamefont {Dmitriev}, \citenamefont {Weng}, \citenamefont {Kippenberg},\ and\ \citenamefont {Bilenko}}]{Voloshin:21}%
  \BibitemOpen
  \bibfield  {author} {\bibinfo {author} {\bibfnamefont {A.~S.}\ \bibnamefont {Voloshin}}, \bibinfo {author} {\bibfnamefont {N.~M.}\ \bibnamefont {Kondratiev}}, \bibinfo {author} {\bibfnamefont {G.~V.}\ \bibnamefont {Lihachev}}, \bibinfo {author} {\bibfnamefont {J.}~\bibnamefont {Liu}}, \bibinfo {author} {\bibfnamefont {V.~E.}\ \bibnamefont {Lobanov}}, \bibinfo {author} {\bibfnamefont {N.~Y.}\ \bibnamefont {Dmitriev}}, \bibinfo {author} {\bibfnamefont {W.}~\bibnamefont {Weng}}, \bibinfo {author} {\bibfnamefont {T.~J.}\ \bibnamefont {Kippenberg}}, \ and\ \bibinfo {author} {\bibfnamefont {I.~A.}\ \bibnamefont {Bilenko}},\ }\href {\doibase 10.1038/s41467-020-20196-y} {\bibfield  {journal} {\bibinfo  {journal} {Nature Communications}\ }\textbf {\bibinfo {volume} {12}},\ \bibinfo {pages} {235} (\bibinfo {year} {2021})}\BibitemShut {NoStop}%
\bibitem [{\citenamefont {Pavlov}\ \emph {et~al.}(2018)\citenamefont {Pavlov}, \citenamefont {Koptyaev}, \citenamefont {Lihachev}, \citenamefont {Voloshin}, \citenamefont {Gorodnitskiy}, \citenamefont {Ryabko}, \citenamefont {Polonsky},\ and\ \citenamefont {Gorodetsky}}]{Pavlov:18}%
  \BibitemOpen
  \bibfield  {author} {\bibinfo {author} {\bibfnamefont {N.~G.}\ \bibnamefont {Pavlov}}, \bibinfo {author} {\bibfnamefont {S.}~\bibnamefont {Koptyaev}}, \bibinfo {author} {\bibfnamefont {G.~V.}\ \bibnamefont {Lihachev}}, \bibinfo {author} {\bibfnamefont {A.~S.}\ \bibnamefont {Voloshin}}, \bibinfo {author} {\bibfnamefont {A.~S.}\ \bibnamefont {Gorodnitskiy}}, \bibinfo {author} {\bibfnamefont {M.~V.}\ \bibnamefont {Ryabko}}, \bibinfo {author} {\bibfnamefont {S.~V.}\ \bibnamefont {Polonsky}}, \ and\ \bibinfo {author} {\bibfnamefont {M.~L.}\ \bibnamefont {Gorodetsky}},\ }\href {\doibase 10.1038/s41566-018-0277-2} {\bibfield  {journal} {\bibinfo  {journal} {Nature Photonics}\ }\textbf {\bibinfo {volume} {12}},\ \bibinfo {pages} {694} (\bibinfo {year} {2018})}\BibitemShut {NoStop}%
\bibitem [{\citenamefont {Yuan}\ \emph {et~al.}(2022)\citenamefont {Yuan}, \citenamefont {Wang}, \citenamefont {Liu}, \citenamefont {Li}, \citenamefont {Shen}, \citenamefont {Gao}, \citenamefont {Chang}, \citenamefont {Jin}, \citenamefont {Feshali}, \citenamefont {Paniccia}, \citenamefont {Bowers},\ and\ \citenamefont {Vahala}}]{Yuan:22}%
  \BibitemOpen
  \bibfield  {author} {\bibinfo {author} {\bibfnamefont {Z.}~\bibnamefont {Yuan}}, \bibinfo {author} {\bibfnamefont {H.}~\bibnamefont {Wang}}, \bibinfo {author} {\bibfnamefont {P.}~\bibnamefont {Liu}}, \bibinfo {author} {\bibfnamefont {B.}~\bibnamefont {Li}}, \bibinfo {author} {\bibfnamefont {B.}~\bibnamefont {Shen}}, \bibinfo {author} {\bibfnamefont {M.}~\bibnamefont {Gao}}, \bibinfo {author} {\bibfnamefont {L.}~\bibnamefont {Chang}}, \bibinfo {author} {\bibfnamefont {W.}~\bibnamefont {Jin}}, \bibinfo {author} {\bibfnamefont {A.}~\bibnamefont {Feshali}}, \bibinfo {author} {\bibfnamefont {M.}~\bibnamefont {Paniccia}}, \bibinfo {author} {\bibfnamefont {J.}~\bibnamefont {Bowers}}, \ and\ \bibinfo {author} {\bibfnamefont {K.}~\bibnamefont {Vahala}},\ }\href {\doibase 10.1364/OE.458109} {\bibfield  {journal} {\bibinfo  {journal} {Opt. Express}\ }\textbf {\bibinfo {volume} {30}},\ \bibinfo {pages} {25147} (\bibinfo {year} {2022})}\BibitemShut {NoStop}%
\bibitem [{\citenamefont {Guo}\ \emph {et~al.}(2022)\citenamefont {Guo}, \citenamefont {McLemore}, \citenamefont {Xiang}, \citenamefont {Lee}, \citenamefont {Wu}, \citenamefont {Jin}, \citenamefont {Kelleher}, \citenamefont {Jin}, \citenamefont {Mason}, \citenamefont {Chang}, \citenamefont {Feshali}, \citenamefont {Paniccia}, \citenamefont {Rakich}, \citenamefont {Vahala}, \citenamefont {Diddams}, \citenamefont {Quinlan},\ and\ \citenamefont {Bowers}}]{Guo:22}%
  \BibitemOpen
  \bibfield  {author} {\bibinfo {author} {\bibfnamefont {J.}~\bibnamefont {Guo}}, \bibinfo {author} {\bibfnamefont {C.~A.}\ \bibnamefont {McLemore}}, \bibinfo {author} {\bibfnamefont {C.}~\bibnamefont {Xiang}}, \bibinfo {author} {\bibfnamefont {D.}~\bibnamefont {Lee}}, \bibinfo {author} {\bibfnamefont {L.}~\bibnamefont {Wu}}, \bibinfo {author} {\bibfnamefont {W.}~\bibnamefont {Jin}}, \bibinfo {author} {\bibfnamefont {M.}~\bibnamefont {Kelleher}}, \bibinfo {author} {\bibfnamefont {N.}~\bibnamefont {Jin}}, \bibinfo {author} {\bibfnamefont {D.}~\bibnamefont {Mason}}, \bibinfo {author} {\bibfnamefont {L.}~\bibnamefont {Chang}}, \bibinfo {author} {\bibfnamefont {A.}~\bibnamefont {Feshali}}, \bibinfo {author} {\bibfnamefont {M.}~\bibnamefont {Paniccia}}, \bibinfo {author} {\bibfnamefont {P.~T.}\ \bibnamefont {Rakich}}, \bibinfo {author} {\bibfnamefont {K.~J.}\ \bibnamefont {Vahala}}, \bibinfo {author} {\bibfnamefont {S.~A.}\ \bibnamefont {Diddams}}, \bibinfo {author} {\bibfnamefont {F.}~\bibnamefont {Quinlan}}, \
  and\ \bibinfo {author} {\bibfnamefont {J.~E.}\ \bibnamefont {Bowers}},\ }\href {\doibase 10.1126/sciadv.abp9006} {\bibfield  {journal} {\bibinfo  {journal} {Science Advances}\ }\textbf {\bibinfo {volume} {8}},\ \bibinfo {pages} {eabp9006} (\bibinfo {year} {2022})}\BibitemShut {NoStop}%
\bibitem [{\citenamefont {Cheng}\ \emph {et~al.}(2023)\citenamefont {Cheng}, \citenamefont {Jin}, \citenamefont {Dai}, \citenamefont {Xiang}, \citenamefont {Guo}, \citenamefont {Zhou}, \citenamefont {Diddams}, \citenamefont {Quinlan}, \citenamefont {Bowers}, \citenamefont {Miller},\ and\ \citenamefont {Rakich}}]{Cheng:23}%
  \BibitemOpen
  \bibfield  {author} {\bibinfo {author} {\bibfnamefont {H.}~\bibnamefont {Cheng}}, \bibinfo {author} {\bibfnamefont {N.}~\bibnamefont {Jin}}, \bibinfo {author} {\bibfnamefont {Z.}~\bibnamefont {Dai}}, \bibinfo {author} {\bibfnamefont {C.}~\bibnamefont {Xiang}}, \bibinfo {author} {\bibfnamefont {J.}~\bibnamefont {Guo}}, \bibinfo {author} {\bibfnamefont {Y.}~\bibnamefont {Zhou}}, \bibinfo {author} {\bibfnamefont {S.~A.}\ \bibnamefont {Diddams}}, \bibinfo {author} {\bibfnamefont {F.}~\bibnamefont {Quinlan}}, \bibinfo {author} {\bibfnamefont {J.}~\bibnamefont {Bowers}}, \bibinfo {author} {\bibfnamefont {O.}~\bibnamefont {Miller}}, \ and\ \bibinfo {author} {\bibfnamefont {P.}~\bibnamefont {Rakich}},\ }\href {\doibase 10.1063/5.0174384} {\bibfield  {journal} {\bibinfo  {journal} {APL Photonics}\ }\textbf {\bibinfo {volume} {8}},\ \bibinfo {pages} {116105} (\bibinfo {year} {2023})}\BibitemShut {NoStop}%
\bibitem [{\citenamefont {Yu}\ \emph {et~al.}(2020)\citenamefont {Yu}, \citenamefont {Gao}, \citenamefont {Ye}, \citenamefont {Chen}, \citenamefont {Sun}, \citenamefont {Xie}, \citenamefont {Srinivasan}, \citenamefont {Zervas}, \citenamefont {Navickaite}, \citenamefont {Geiselmann},\ and\ \citenamefont {Beling}}]{Yu:20}%
  \BibitemOpen
  \bibfield  {author} {\bibinfo {author} {\bibfnamefont {Q.}~\bibnamefont {Yu}}, \bibinfo {author} {\bibfnamefont {J.}~\bibnamefont {Gao}}, \bibinfo {author} {\bibfnamefont {N.}~\bibnamefont {Ye}}, \bibinfo {author} {\bibfnamefont {B.}~\bibnamefont {Chen}}, \bibinfo {author} {\bibfnamefont {K.}~\bibnamefont {Sun}}, \bibinfo {author} {\bibfnamefont {L.}~\bibnamefont {Xie}}, \bibinfo {author} {\bibfnamefont {K.}~\bibnamefont {Srinivasan}}, \bibinfo {author} {\bibfnamefont {M.}~\bibnamefont {Zervas}}, \bibinfo {author} {\bibfnamefont {G.}~\bibnamefont {Navickaite}}, \bibinfo {author} {\bibfnamefont {M.}~\bibnamefont {Geiselmann}}, \ and\ \bibinfo {author} {\bibfnamefont {A.}~\bibnamefont {Beling}},\ }\href {\doibase 10.1364/OE.387939} {\bibfield  {journal} {\bibinfo  {journal} {Opt. Express}\ }\textbf {\bibinfo {volume} {28}},\ \bibinfo {pages} {14824} (\bibinfo {year} {2020})}\BibitemShut {NoStop}%
\bibitem [{\citenamefont {Xiang}\ \emph {et~al.}(2021)\citenamefont {Xiang}, \citenamefont {Liu}, \citenamefont {Guo}, \citenamefont {Chang}, \citenamefont {Wang}, \citenamefont {Weng}, \citenamefont {Peters}, \citenamefont {Xie}, \citenamefont {Zhang}, \citenamefont {Riemensberger}, \citenamefont {Selvidge}, \citenamefont {Kippenberg},\ and\ \citenamefont {Bowers}}]{Xiang:21}%
  \BibitemOpen
  \bibfield  {author} {\bibinfo {author} {\bibfnamefont {C.}~\bibnamefont {Xiang}}, \bibinfo {author} {\bibfnamefont {J.}~\bibnamefont {Liu}}, \bibinfo {author} {\bibfnamefont {J.}~\bibnamefont {Guo}}, \bibinfo {author} {\bibfnamefont {L.}~\bibnamefont {Chang}}, \bibinfo {author} {\bibfnamefont {R.~N.}\ \bibnamefont {Wang}}, \bibinfo {author} {\bibfnamefont {W.}~\bibnamefont {Weng}}, \bibinfo {author} {\bibfnamefont {J.}~\bibnamefont {Peters}}, \bibinfo {author} {\bibfnamefont {W.}~\bibnamefont {Xie}}, \bibinfo {author} {\bibfnamefont {Z.}~\bibnamefont {Zhang}}, \bibinfo {author} {\bibfnamefont {J.}~\bibnamefont {Riemensberger}}, \bibinfo {author} {\bibfnamefont {J.}~\bibnamefont {Selvidge}}, \bibinfo {author} {\bibfnamefont {T.~J.}\ \bibnamefont {Kippenberg}}, \ and\ \bibinfo {author} {\bibfnamefont {J.~E.}\ \bibnamefont {Bowers}},\ }\href {\doibase 10.1126/science.abh2076} {\bibfield  {journal} {\bibinfo  {journal} {Science}\ }\textbf {\bibinfo {volume} {373}},\ \bibinfo {pages} {99} (\bibinfo {year}
  {2021})}\BibitemShut {NoStop}%
\bibitem [{\citenamefont {Liu}\ \emph {et~al.}(2020{\natexlab{b}})\citenamefont {Liu}, \citenamefont {Tian}, \citenamefont {Lucas}, \citenamefont {Raja}, \citenamefont {Lihachev}, \citenamefont {Wang}, \citenamefont {He}, \citenamefont {Liu}, \citenamefont {Anderson}, \citenamefont {Weng}, \citenamefont {Bhave},\ and\ \citenamefont {Kippenberg}}]{Liu:20a}%
  \BibitemOpen
  \bibfield  {author} {\bibinfo {author} {\bibfnamefont {J.}~\bibnamefont {Liu}}, \bibinfo {author} {\bibfnamefont {H.}~\bibnamefont {Tian}}, \bibinfo {author} {\bibfnamefont {E.}~\bibnamefont {Lucas}}, \bibinfo {author} {\bibfnamefont {A.~S.}\ \bibnamefont {Raja}}, \bibinfo {author} {\bibfnamefont {G.}~\bibnamefont {Lihachev}}, \bibinfo {author} {\bibfnamefont {R.~N.}\ \bibnamefont {Wang}}, \bibinfo {author} {\bibfnamefont {J.}~\bibnamefont {He}}, \bibinfo {author} {\bibfnamefont {T.}~\bibnamefont {Liu}}, \bibinfo {author} {\bibfnamefont {M.~H.}\ \bibnamefont {Anderson}}, \bibinfo {author} {\bibfnamefont {W.}~\bibnamefont {Weng}}, \bibinfo {author} {\bibfnamefont {S.~A.}\ \bibnamefont {Bhave}}, \ and\ \bibinfo {author} {\bibfnamefont {T.~J.}\ \bibnamefont {Kippenberg}},\ }\href {\doibase 10.1038/s41586-020-2465-8} {\bibfield  {journal} {\bibinfo  {journal} {Nature}\ }\textbf {\bibinfo {volume} {583}},\ \bibinfo {pages} {385} (\bibinfo {year} {2020}{\natexlab{b}})}\BibitemShut {NoStop}%
\bibitem [{\citenamefont {Snigirev}\ \emph {et~al.}(2023)\citenamefont {Snigirev}, \citenamefont {Riedhauser}, \citenamefont {Lihachev}, \citenamefont {Churaev}, \citenamefont {Riemensberger}, \citenamefont {Wang}, \citenamefont {Siddharth}, \citenamefont {Huang}, \citenamefont {M{\"o}hl}, \citenamefont {Popoff}, \citenamefont {Drechsler}, \citenamefont {Caimi}, \citenamefont {H{\"o}nl}, \citenamefont {Liu}, \citenamefont {Seidler},\ and\ \citenamefont {Kippenberg}}]{Snigirev:23}%
  \BibitemOpen
  \bibfield  {author} {\bibinfo {author} {\bibfnamefont {V.}~\bibnamefont {Snigirev}}, \bibinfo {author} {\bibfnamefont {A.}~\bibnamefont {Riedhauser}}, \bibinfo {author} {\bibfnamefont {G.}~\bibnamefont {Lihachev}}, \bibinfo {author} {\bibfnamefont {M.}~\bibnamefont {Churaev}}, \bibinfo {author} {\bibfnamefont {J.}~\bibnamefont {Riemensberger}}, \bibinfo {author} {\bibfnamefont {R.~N.}\ \bibnamefont {Wang}}, \bibinfo {author} {\bibfnamefont {A.}~\bibnamefont {Siddharth}}, \bibinfo {author} {\bibfnamefont {G.}~\bibnamefont {Huang}}, \bibinfo {author} {\bibfnamefont {C.}~\bibnamefont {M{\"o}hl}}, \bibinfo {author} {\bibfnamefont {Y.}~\bibnamefont {Popoff}}, \bibinfo {author} {\bibfnamefont {U.}~\bibnamefont {Drechsler}}, \bibinfo {author} {\bibfnamefont {D.}~\bibnamefont {Caimi}}, \bibinfo {author} {\bibfnamefont {S.}~\bibnamefont {H{\"o}nl}}, \bibinfo {author} {\bibfnamefont {J.}~\bibnamefont {Liu}}, \bibinfo {author} {\bibfnamefont {P.}~\bibnamefont {Seidler}}, \ and\ \bibinfo {author} {\bibfnamefont
  {T.~J.}\ \bibnamefont {Kippenberg}},\ }\href {\doibase 10.1038/s41586-023-05724-2} {\bibfield  {journal} {\bibinfo  {journal} {Nature}\ }\textbf {\bibinfo {volume} {615}},\ \bibinfo {pages} {411} (\bibinfo {year} {2023})}\BibitemShut {NoStop}%
\bibitem [{\citenamefont {Gorodetsky}\ \emph {et~al.}(2000)\citenamefont {Gorodetsky}, \citenamefont {Pryamikov},\ and\ \citenamefont {Ilchenko}}]{Gorodetsky:00}%
  \BibitemOpen
  \bibfield  {author} {\bibinfo {author} {\bibfnamefont {M.~L.}\ \bibnamefont {Gorodetsky}}, \bibinfo {author} {\bibfnamefont {A.~D.}\ \bibnamefont {Pryamikov}}, \ and\ \bibinfo {author} {\bibfnamefont {V.~S.}\ \bibnamefont {Ilchenko}},\ }\href {\doibase 10.1364/JOSAB.17.001051} {\bibfield  {journal} {\bibinfo  {journal} {J. Opt. Soc. Am. B}\ }\textbf {\bibinfo {volume} {17}},\ \bibinfo {pages} {1051} (\bibinfo {year} {2000})}\BibitemShut {NoStop}%
\bibitem [{\citenamefont {Li}\ \emph {et~al.}(2013{\natexlab{b}})\citenamefont {Li}, \citenamefont {Eftekhar}, \citenamefont {Xia},\ and\ \citenamefont {Adibi}}]{Li:13}%
  \BibitemOpen
  \bibfield  {author} {\bibinfo {author} {\bibfnamefont {Q.}~\bibnamefont {Li}}, \bibinfo {author} {\bibfnamefont {A.~A.}\ \bibnamefont {Eftekhar}}, \bibinfo {author} {\bibfnamefont {Z.}~\bibnamefont {Xia}}, \ and\ \bibinfo {author} {\bibfnamefont {A.}~\bibnamefont {Adibi}},\ }\href {\doibase 10.1103/PhysRevA.88.033816} {\bibfield  {journal} {\bibinfo  {journal} {Phys. Rev. A}\ }\textbf {\bibinfo {volume} {88}},\ \bibinfo {pages} {033816} (\bibinfo {year} {2013}{\natexlab{b}})}\BibitemShut {NoStop}%
\bibitem [{\citenamefont {Cai}\ \emph {et~al.}(2000)\citenamefont {Cai}, \citenamefont {Painter},\ and\ \citenamefont {Vahala}}]{Cai:00}%
  \BibitemOpen
  \bibfield  {author} {\bibinfo {author} {\bibfnamefont {M.}~\bibnamefont {Cai}}, \bibinfo {author} {\bibfnamefont {O.}~\bibnamefont {Painter}}, \ and\ \bibinfo {author} {\bibfnamefont {K.~J.}\ \bibnamefont {Vahala}},\ }\href {\doibase 10.1103/PhysRevLett.85.74} {\bibfield  {journal} {\bibinfo  {journal} {Phys. Rev. Lett.}\ }\textbf {\bibinfo {volume} {85}},\ \bibinfo {pages} {74} (\bibinfo {year} {2000})}\BibitemShut {NoStop}%
\end{thebibliography}%


\begin{thebibliography}{1}
\expandafter\ifx\csname url\endcsname\relax
  \def\url#1{\texttt{#1}}\fi
\expandafter\ifx\csname urlprefix\endcsname\relax\def\urlprefix{URL }\fi
\providecommand{\bibinfo}[2]{#2}
\providecommand{\eprint}[2][]{\url{#2}}

\bibitem{Ye:23}
\bibinfo{author}{Ye, Z.} \emph{et~al.}
\newblock \bibinfo{title}{Foundry manufacturing of tight-confinement, dispersion-engineered, ultralow-loss silicon nitride photonic integrated circuits}.
\newblock \emph{\bibinfo{journal}{Photon. Res.}} \textbf{\bibinfo{volume}{11}}, \bibinfo{pages}{558--568} (\bibinfo{year}{2023}).
\newblock \urlprefix\url{https://opg.optica.org/prj/abstract.cfm?URI=prj-11-4-558}.

\bibitem{Luo:23}
\bibinfo{author}{Luo, Y.-H.} \emph{et~al.}
\newblock \bibinfo{title}{A wideband, high-resolution vector spectrum analyzer for integrated photonics}.
\newblock \emph{\bibinfo{journal}{arXiv}} \textbf{\bibinfo{volume}{2304.04295}} (\bibinfo{year}{2023}).

\bibitem{Kondratiev:17}
\bibinfo{author}{Kondratiev, N.~M.} \emph{et~al.}
\newblock \bibinfo{title}{Self-injection locking of a laser diode to a high-{Q} {WGM} microresonator}.
\newblock \emph{\bibinfo{journal}{Optics Express}} \textbf{\bibinfo{volume}{25}}, \bibinfo{pages}{28167--28178} (\bibinfo{year}{2017}).

\bibitem{Kondratiev:23}
\bibinfo{author}{Kondratiev, N.~M.} \emph{et~al.}
\newblock \bibinfo{title}{Recent advances in laser self-injection locking to high-q microresonators}.
\newblock \emph{\bibinfo{journal}{Frontiers of Physics}} \textbf{\bibinfo{volume}{18}}, \bibinfo{pages}{21305} (\bibinfo{year}{2023}).
\newblock \urlprefix\url{https://doi.org/10.1007/s11467-022-1245-3}.

\bibitem{Coldren:12}
\bibinfo{author}{Coldren, L.}, \bibinfo{author}{Corzine, S.} \& \bibinfo{author}{Mashanovitch, M.}
\newblock \emph{\bibinfo{title}{Diode Lasers and Photonic Integrated Circuits}}.
\newblock Wiley Series in Microwave and Optical Engineering (\bibinfo{publisher}{Wiley}, \bibinfo{year}{2012}).
\newblock \urlprefix\url{https://books.google.ch/books?id=GBB1kOYONT4C}.

\bibitem{Okoshi:88}
\bibinfo{author}{Okoshi, T.} \& \bibinfo{author}{Kikuchi, K.}
\newblock \emph{\bibinfo{title}{Coherent optical fiber communications}}, vol.~\bibinfo{volume}{4} (\bibinfo{publisher}{Springer Science \& Business Media}, \bibinfo{year}{1988}).

\bibitem{camatel:08}
\bibinfo{author}{Camatel, S.} \& \bibinfo{author}{Ferrero, V.}
\newblock \bibinfo{title}{Narrow linewidth cw laser phase noise characterization methods for coherent transmission system applications}.
\newblock \emph{\bibinfo{journal}{Journal of Lightwave Technology}} \textbf{\bibinfo{volume}{26}}, \bibinfo{pages}{3048--3055} (\bibinfo{year}{2008}).
\newblock \urlprefix\url{https://ieeexplore.ieee.org/document/4738475}.

\bibitem{Jin:21}
\bibinfo{author}{Jin, W.} \emph{et~al.}
\newblock \bibinfo{title}{Hertz-linewidth semiconductor lasers using cmos-ready ultra-high-q microresonators}.
\newblock \emph{\bibinfo{journal}{Nature Photonics}} \textbf{\bibinfo{volume}{15}}, \bibinfo{pages}{346--353} (\bibinfo{year}{2021}).
\newblock \urlprefix\url{https://doi.org/10.1038/s41566-021-00761-7}.

\end{thebibliography}
\end{document}

\title{Supplementary Information to: A chip-integrated comb-based microwave oscillator}

\author{Wei Sun}
\thanks{These authors contributed equally to this work.}
\affiliation{International Quantum Academy, Shenzhen 518048, China}

\author{Zhiyang Chen}
\thanks{These authors contributed equally to this work.}
\affiliation{International Quantum Academy, Shenzhen 518048, China}
\affiliation{Shenzhen Institute for Quantum Science and Engineering, Southern University of Science and Technology, Shenzhen 518055, China}

\author{Linze Li}
\thanks{These authors contributed equally to this work.}
\affiliation{School of Information Science and Technology, ShanghaiTech University, Shanghai 201210, China}

\author{Chen Shen}
\thanks{These authors contributed equally to this work.}
\affiliation{International Quantum Academy, Shenzhen 518048, China}
\affiliation{Qaleido Photonics, Hangzhou 310000, China}

\author{Jinbao Long}
\affiliation{International Quantum Academy, Shenzhen 518048, China}

\author{Huamin Zheng}
\affiliation{International Quantum Academy, Shenzhen 518048, China}
\affiliation{College of Electronics and Information Engineering, Shenzhen University, Shenzhen 518000, China}

\author{Luyu Wang}
\affiliation{School of Information Science and Technology, ShanghaiTech University, Shanghai 201210, China}

\author{Qiushi Chen}
\affiliation{School of Information Science and Technology, ShanghaiTech University, Shanghai 201210, China}

\author{Zhouze Zhang}
\affiliation{School of Information Science and Technology, ShanghaiTech University, Shanghai 201210, China}

\author{Baoqi Shi}
\affiliation{International Quantum Academy, Shenzhen 518048, China}
\affiliation{Department of Optics and Optical Engineering, University of Science and Technology of China, Hefei 230026, China}

\author{Shichang Li}
\affiliation{International Quantum Academy, Shenzhen 518048, China}
\affiliation{Shenzhen Institute for Quantum Science and Engineering, Southern University of Science and Technology,
Shenzhen 518055, China}

\author{Lan Gao}
\affiliation{International Quantum Academy, Shenzhen 518048, China}
\affiliation{Shenzhen Institute for Quantum Science and Engineering, Southern University of Science and Technology,
Shenzhen 518055, China}

\author{Yi-Han Luo}
\affiliation{International Quantum Academy, Shenzhen 518048, China}

\author{Baile Chen}
\email[]{chenbl@shanghaitech.edu.cn}
\affiliation{School of Information Science and Technology, ShanghaiTech University, Shanghai 201210, China}

\author{Junqiu Liu}
\email[]{liujq@iqasz.cn}
\affiliation{International Quantum Academy, Shenzhen 518048, China}
\affiliation{Hefei National Laboratory, University of Science and Technology of China, Hefei 230088, China}

\maketitle
%\pagebreak

%%%%%%%%%%%%%%%%%%%%%%%%%%%%%%%%%%%%%%%%%%%%%%%%%%%%%%%%%%%%%%%

\section{Fabrication process of Si$_3$N$_4$ chip}
The high-$Q$ Si$_3$N$_4$ microresonator chips are fabricated using an optimized DUV subtractive process on 6-inch wafers \cite{Ye:23}. 
The process flow is shown in Supplementary Fig.~\ref{FigS:SiN}. 
First, an LPCVD Si$_3$N$_4$ film is deposited on a clean thermal wet SiO$_2$ substrate.
Compared to thick Si$_3$N$_4$ films suffering from cracks due to intrinsic tensile stress, the 300-nm-thick Si$_3$N$_4$ films do not exhibit any cracks during the fabrication. 
A SiO$_2$ film is further deposited as an etch hardmask.  
Afterwards, DUV stepper photolithography is performed, followed by dry etching to transfer the pattern from the photoresist to the SiO$_2$ hardmask and then to the Si$_3$N$_4$ layer.
In the dry etching, etchants of CHF$_3$ and O$_2$ are used to create ultra-smooth and vertical etched surfaces, which are critical for low optical loss in waveguides. 
Then the photoresist is removed and a thermal annealing is applied in the nitrogen atmosphere at 1200$^\circ$C to drive out hydrogen contents that cause optical absorption loss in the waveguides.
Then 3-$\mu$m-thick SiO$_2$ top cladding is deposited on the wafer and thermally annealed again at 1200$^\circ$C to drive out hydrogen contents.
Afterwards, platinum heaters are deposited on the substrate via an electron-beam evaporator, and patterned via a lift-off process. 
The metal heaters have negligible impact on the optical loss of the Si$_3$N$_4$ waveguide due to the thick SiO$_2$ cladding.
Contact UV photolithography and additional deep dry etching are performed to create smooth chip facets and define the chip size, critical for later hybrid integration and packaging.
Finally, the wafer is separated into individual chips by backside grinding.

\begin{figure*}[h!]
\renewcommand{\figurename}{Supplementary Figure}
\centering
\includegraphics{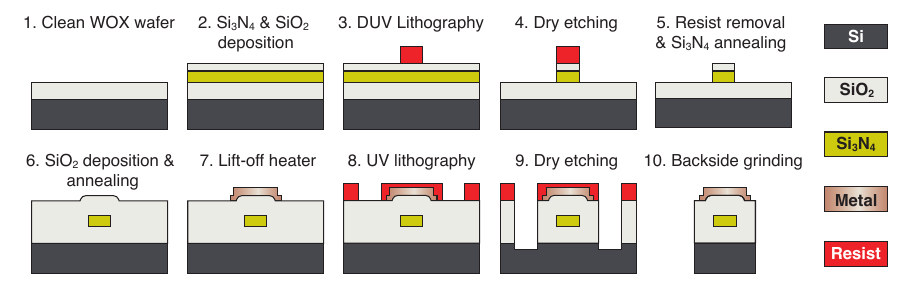}
\caption{
\textbf{The DUV subtractive process flow of the 6-inch-wafer Si$_3$N$_4$ foundry fabrication.}
WOX, thermal wet oxide (SiO$_2$).
}
\label{FigS:SiN}
\end{figure*}

\section{Loss measurement over large bandwidth}
By fitting the transmission spectrum of a resonance mode, the intrinsic loss $\kappa_0/2\pi$ and the external coupling rate $\kappa_\text{ex}/2\pi$ can be extracted, as shown in Supplementary Fig.~\ref{FigS:Loss}a.
The resonance mode locates at frequency $\nu = \omega/2\pi =$ 200.5 THz.
The intrinsic loss is $\kappa_0/2\pi =$ 8.6 MHz, and the external coupling rate is $\kappa_\text{ex}/2\pi =$ 4.7 MHz, which indicates an under coupling of the resonance.
Scanning a laser from $\nu =$ 202.6 THz to 182.8 THz (wavelength from 1480 to 1640 nm), the loss over a large bandwidth can be traced, as shown in Supplementary Fig.~\ref{FigS:Loss}b.
The overall measurements are performed using a vector spectrum analyser\cite{Luo:23}.
\begin{figure*}[h!]
\renewcommand{\figurename}{Supplementary Figure}
\centering
\includegraphics{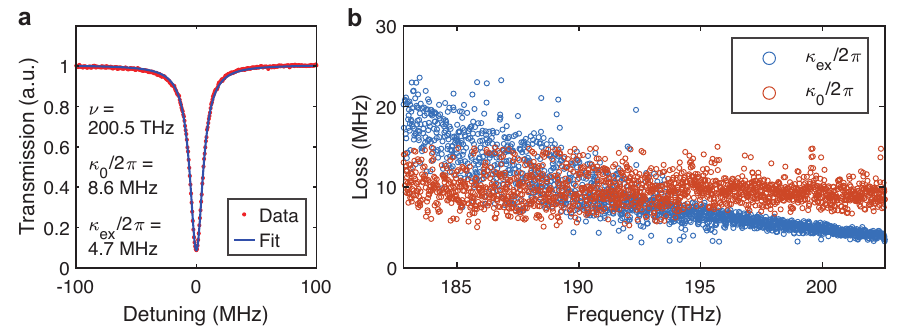}
\caption{
\textbf{Intrinsic and external coupling loss of the Si$_3$N$_4$ microresonator.}
\textbf{a}. Resonance fitting at frequency $\nu =$ 200.5 THz.
Red dots are measured data.
Blue line is the corresponding fit.
\textbf{b}. Loss measurement over 20 THz.
The red circles are data of intrinsic loss.
The blue circles are data of external coupling loss.
}
\label{FigS:Loss}
\end{figure*}

\section{Fabrication process of PD chip}

The epitaxial structure of the PD chip is grown by metal-organic chemical vapor deposition (MOCVD) on a 2-inch semi-insulating InP wafer. 
The fabrication process is shown in Supplementary Fig.~\ref{FigS:PD}. 
The fabrication process begins with the metal deposition of the p-contacts (step 2), composed of Ti/Pt/Au layers (20/30/120 nm). 
Then, a tipple-mesa structure is defined by a combination of dry and wet etch process (step 3). 
Inductively coupled plasma (ICP) dry etching is performed to achieve vertical sidewalls and precise control of the first p-mesa and the second waveguide mesa. 
The third n-mesa is terminated at the semi-insulating InP substrate by wet chemical etching (H$_3$PO$_4$:HCl = 1:3) to ensure electrical isolation. 
After the deposition of GeAu/Ni/Au layers (40/20/120 nm) as the n-contact metals (step 4), rapid thermal annealing process at 360$^\circ$C is conducted for lower contact resistance. 
A benzocyclobutene (BCB) layer is implemented to passivate the sidewall of p-mesa and provide stable support for the subsequent electrodes (step 5). 
Finally, the p-mesa is connected to coplanar waveguide (CPWs) pads of 50 $\Omega$ characteristic impedance by metal deposition (step 6). 
Such a non-suspended structure eliminates the necessity for air-bridge structures, consequently ensuring a consistent and stable connection between p-mesa and CPWs.

\begin{figure*}[h!]
\renewcommand{\figurename}{Supplementary Figure}
\centering
\includegraphics{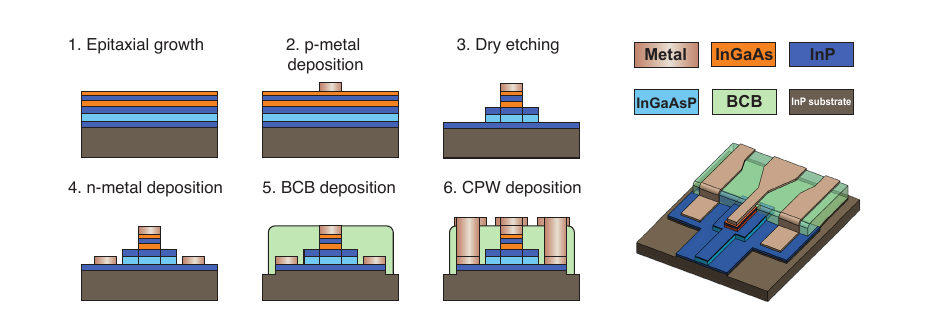}
\caption{
\textbf{Fabrication process of the photodetector chip.}
BCB, benzocyclobutene.
CPW, coplanar waveguide.
}
\label{FigS:PD}
\end{figure*}

\section{DFB laser characterization}
The performance of the DFB laser is characterized as shown in Supplementary Fig.~\ref{FigS:DFB}. 
The DFB can emit 160 mW laser at 500 mA current with the laser threshold of 55 mA.
In the 500 mA current range, the DFB laser's emission wavelength can be tuned over 1 nm from the starting wavelength 1548.786 nm at the temperature of 30$^\circ$C.
In the edge-coupling scheme, the coupling efficiency is measured at different currents from 280 mA to 480 mA, with a mean coupling efficiency of $\sim23\%$.

\begin{figure*}[h!]
\renewcommand{\figurename}{Supplementary Figure}
\centering
\includegraphics{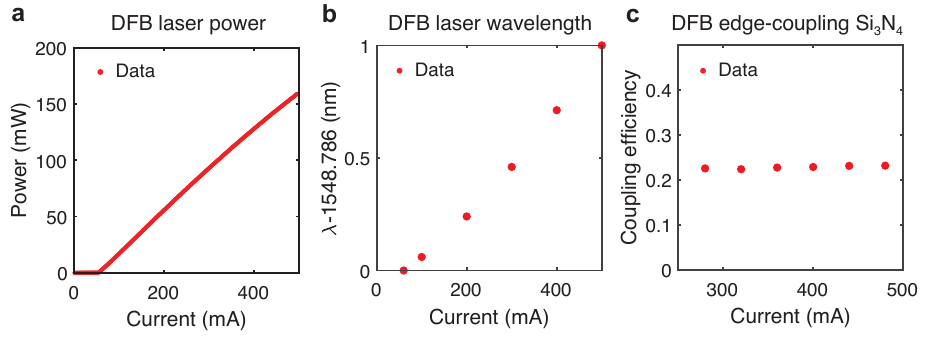}
\caption{
\textbf{DFB laser characterization.}
\textbf{a}. 
Measured DFB output power versus the applied current.
\textbf{b}. 
Measured DFB laser frequency versus the applied current.
\textbf{c}. 
Measured coupling efficiency between the DFB laser chip and the Si$_3$N$_4$ chip.
}
\label{FigS:DFB}
\end{figure*}

\pagebreak
\section{Self-injection locking to the microresonator}
A comprehensive theory on self-injection locking of a laser diode to a microresonator is presented in Ref. \cite{Kondratiev:17, Kondratiev:23}.
By a simple model and reasonable approximation, the stable solution of the combined system is described as
\begin{equation}
    \xi = \zeta +  \frac{K}{2} \frac{2\zeta \cos \bar{ \psi} +  \left ( 1 +  \beta ^2 - \zeta^2 \right ) \sin \bar{ \psi} }{\left ( 1 +  \beta ^2 - \zeta^2 \right )^2 +  4\zeta ^2}. 
    \label{Eq_SIL1}
\end{equation}
where $\xi$ represents the frequency detuning between the generated light of the system and the resonance of the microresonator, 
$\zeta$ represents the frequency detuning between the free-running laser and the resonance of the microresonator, 
$K=8\eta \beta \sqrt{1+\alpha_{\text{g}}^2} \kappa_{\text{do}} /(R_\text{o}\kappa)$ denotes the combined coupling coefficient of the system,
$\bar{ \psi}=\psi_0 + \kappa \tau_\text{s}\zeta/2$ is the phase delay between the laser and the microresonator, and $\beta$ is the back-scattering-related mode splitting of the microresonator.
Parameters $\xi$, $\zeta$ and $\beta$ are in the unit of $\kappa/2$ where $\kappa /2\pi$ is the loaded linewidth of the Si$_3$N$_4$ microresonator.

Experimentally, $\kappa$ is measured as $2\pi \times 17.6$ MHz. 
The intrinsic loss rate $\kappa_o$ and the external coupling rate $\kappa _\text{ext}$ are $2\pi \times 11.9$ MHz and $2\pi \times 6.7$ MHz, respectively. 
The coupling efficiency $\eta = \kappa_\text{ex}/\kappa$ of the microresonator is 0.36. 
The mode splitting of the resonance $\beta$ is nearly negligible for high-Q microresonator. 
The phase-amplitude coupling factor $\alpha_\text{g}$ in the DFB laser is set as 2.5 \cite{Kondratiev:17}. 
The front reflection $R_\text{o}$ and the rear reflection $R_\text{e}$ of the DFB laser are 0.1\% and 97\% provided by the manufacturer, respectively.
$\kappa_\text{do}$ is the output beam coupling rate.
Considering the DFB laser loses most power from the output, the quality factor of the DFB cavity $Q_\text{d}$ can be represented as $Q_\text{d}=\omega/\kappa_\text{do}$, where $\omega/2\pi$ is the DFB laser frequency.
The time delay $\tau_\text{s}$ between the DFB laser and the microresonator is estimated as $3\times 10^{-11}$ s. 

In case of $\kappa \tau_\text{s} \ll 1$, $\beta \ll 1$, which is true in our experiment, and $\psi_0=0$ for convenience, Eq.~\ref{Eq_SIL1} can be re-written as 
\begin{equation}
    \xi = \zeta +  K \frac{\zeta}{\left ( 1+ \zeta^2 \right )^2 }. 
    \label{Eq_SIL2}
\end{equation}
For sufficiently large $K$ ($K \gg 4$), the locking range in the SIL region is $\Delta\omega \approx 3\sqrt{3}K\kappa/16 \approx 2\omega\sqrt{1+\alpha_\text{g}^2}\eta \beta/(Q_\text{d}R_\text{o})$ by solving $\partial \xi /\partial \zeta = 0$.
As measured in the main text, the locking range is about $\Delta \omega/2\pi\approx 4$ GHz, and $K$ is estimated as 700.
$\beta$ is set to 0.004 for calculation.
Applying the above parameters in Eq.~\ref{Eq_SIL1}, the solution is shown in Fig.~3a in the main text.

\vspace{1cm}
\section{Self-injection locking to the waveguide Fabry-P\'erot cavity}
The DFB laser's frequency can be locked to the Si$_3$N$_4$ bus waveguide with edge-coupling to the microresonator. 
In this case, the bus waveguide behaves as a Fabry-P\'erot (FP) cavity. 
The locked laser frequency is measured by the beat-note frequency with the reference frequency. 
The length of the DFB and the Si$_3$N$_4$ FP cavity are 1.5 mm and 5.0 mm, respectively. 
The forward and backward dynamics of laser frequency tuning are presented in Supplementary Fig.~\ref{Fig:MP}b.
The quarter-wave shifted DFB laser is simulated by the transmission matrix theory of gratings. 
The effective refractive index of the active grating is set to $n_\text{e}=3.3$ with grating couple efficiency $\kappa_\text{g} = 10.3$ cm$^{-1}$. 
The effective refractive index of the Si$_3$N$_4$ waveguide is set to $n_\text{p} = 1.9$. 
The laser frequency is linearly dependent on the DFB current in free-running state, as the green lines in Supplementary Fig.~\ref{Fig:MP}b. 
The effective reflectivity for the hybrid system is defined by
\cite{Coldren:12}
\begin{equation}
    r_{\text{eff}} = r_{\text{g}}(n_{\text{e}}, \kappa_\text{g} ) +\frac{t_{\text{g}}^{2}(n_{\text{e}}, \kappa_\text{g} )\cdot r_{\text{p}}e^{-2j\beta_{\text{p}}L_{\text{p}}}} {1+r_{\text{g}}(n_{\text{e}}, \kappa_\text{g} )\cdot r_{\text{p}}e^{-2j\beta_{\text{p}}L_{\text{p}}}},
\end{equation}
where $r_{\text{g}}$ and $t_{\text{g}}$ are the effective reflection and transmission coefficients for the front DFB grating, respectively. 
And $r_{\text{p}}$ is the reflection coefficient for the interface of the Si$_3$N$_4$ waveguide, $\beta_{\text{p}}=n_{\text{p}}k$ where $k$ is the wave vector. 
$L_{\text{p}}$ is the length of the Si$_3$N$_4$ waveguide.
The round-trip light field in the cavity is
\begin{equation}
    E_{\text{rp}} = r_{\text{r}}(n_{\text{e}}, \kappa_\text{g} )\cdot r_{\text{eff}} e^{-2j\beta_{\text{a}}L_{\text{a}}},
\end{equation}
where $r_{\text{r}}$ is the effective reflection coefficient for the rear DFB grating, $\beta_{\text{a}}=n_{\text{e}}k$, 
and $L_{\text{a}}$ is the length of DFB.
The laser frequency is obtained by the cavity round-trip phase condition.
Through forward and backward current tuning, the laser frequency dynamics is presented as blue and red lines in Supplementary Fig.~\ref{Fig:MP}b. 
The FSR of the Si$_3$N$_4$ FP cavity about 15.9 GHz indicates that the laser mode hopping is due to the coupling of the DFB mode and the FP mode. 
The hysteresis loop can change with different phases between the DFB laser and the Si$_3$N$_4$ waveguide. 
The simulation with $0.1\pi$ phase is shown in Supplementary Fig.~\ref{Fig:MP}b lower panel.
\begin{figure*}[t!]
\renewcommand{\figurename}{Supplementary Figure}
\centering
\includegraphics{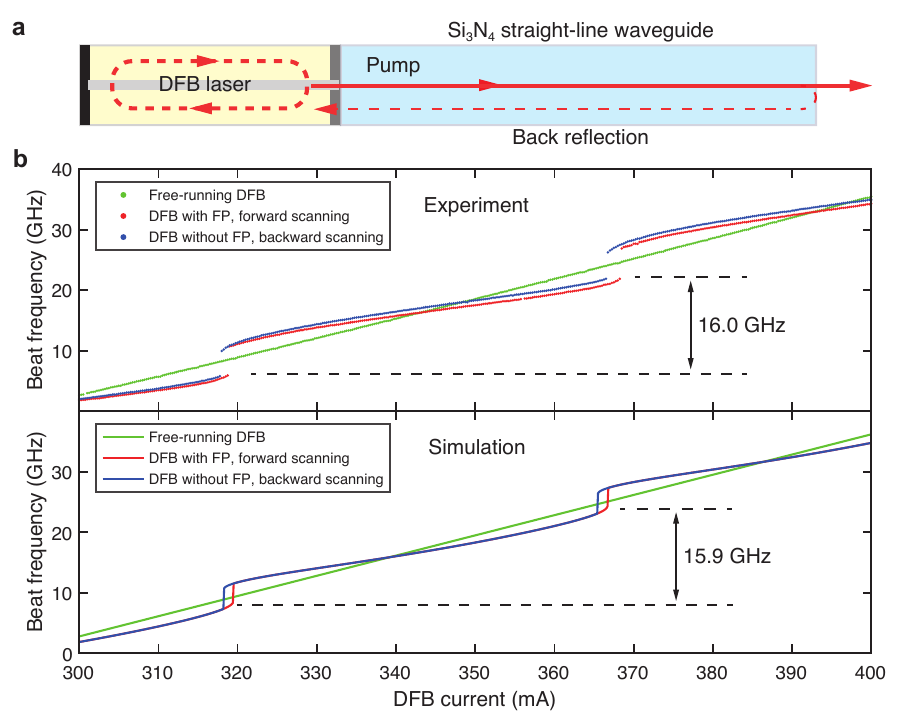}
\caption{
\textbf{Laser self-injection locking to the Si$_3$N$_4$ bus waveguide.}
\textbf{a}. 
Schematic of the DFB laser locked to an FP cavity.
\textbf{b}. 
Hysteresis behavior when scanning the DFB laser frequency via the DFB current. 
When the current is tuned from 300 to 400 mA, the frequency mode hopping is observed at the FP cavity's resonant frequency (red curve). 
While in the backward tuning, the mode hopping is also observed despite that the hopping current point is not the same as that in the forward scanning. 
The blue curve shows a slight current offset. 
The green line is the frequency tuning of the free-running DFB.
The simulation is presented in the lower panel using the transmission matrix method and coupled-mode theory. 
The mode hopping period of about 15.9 GHz (experiment about 16.0 GHz) corresponds to the FSR of the Si$_3$N$_4$ FP cavity.
}
\label{Fig:MP}
\end{figure*}

\pagebreak
\section{Frequency noise measurement}
The laser frequency noise is measured by a delayed self-heterodyne interferometer (DSHI)\cite{Okoshi:88}. 
The experimental setup is shown in Supplementary Fig.~\ref{FigS:selfH}.
The laser passes a polarization controller (PC) with its polarization aligned to the slow axis of the polarization-maintained fiber, which ensures the largest diffraction efficiency of the followed acousto-optic modulator (AOM).
The modulation frequency on the AOM is 80 MHz.
Parallel to the path of the light through the AOM, the other branch is sent into a 2-km-long fiber.
The time delay $\tau_{\text{d}}$ between the two branches is estimated far less than the coherent time $\tau_{\text{c}}$ of the laser, i.e. $\tau_{\text{d}} \ll \tau_{\text{c}}$, to ensure a sub-coherent measurement.
A fiber coupler combines the lights in both branches, with a PC in one path to ensure the same polarization in the fiber coupler.
The beat signal of the two branches is detected by a photodetector (PD, Finisar XPDV3120R). 
The power spectral density (PSD) of the phase noise $S_{\Delta \phi} (f)$ is measured by a commercial phase noise analyser (PNA, Rohde \& Schwarz FSWP50).
The PSD of the frequency noise $S_{\nu} (f)$ can be directly derived form $S_{\Delta \phi} (f)$ by \cite{camatel:08}
\begin{equation}
    S_{\nu} (f)=\frac{f^2}{4\left [ \sin \left ( \pi f \tau_{\text{d}}  \right )  \right ]^2 } S_{\Delta \phi} (f),
\end{equation}
where $f$ is the Fourier frequency offset, $\nu$ is the laser frequency.
The intrinsic linewidth $\Delta \nu_{\text{intrinsic}}$ of the laser can be obtained from the white frequency noise in the high Fourier frequency offset by $\Delta \nu_{\text{intrinsic}} = 2\pi S_{\nu}^0$, where $S_{\nu}^0$ is the single-sideband PSD of the white frequency noise\cite{Jin:21,camatel:08}.

\begin{figure*}[h!]
\renewcommand{\figurename}{Supplementary Figure}
\centering
\includegraphics{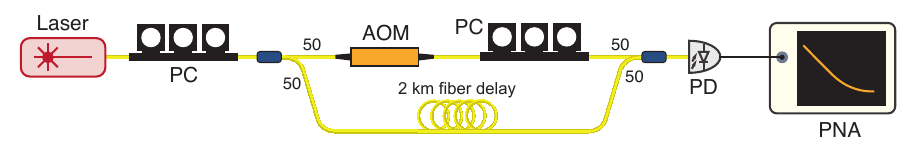}
\caption{
\textbf{Experimental setup of delayed self-heterodyne interferometer.}
PC, polarization controller.
AOM, acousto-optic modulator with a center modulation frequency of 80 MHz.
PD, photodetecter, Finisar XPDV3120R.
PNA, phase noise analyser, R\&D FSWP50.
}
\label{FigS:selfH}
\end{figure*}

%%%%%%%%%%%%%%%%%%%%%%%%%%%%%%%%%%%%%%%%%%%
% \pagebreak

%%%%%%%%%%%%%%%%%%%%%%%%%%%%%%%%%%%%%%%%%%%

\section*{Supplementary References}
\bigskip
\bibliographystyle{naturemag}
\bibliography{bibliography}